\documentclass[12pt]{report}
\usepackage[letterpaper]{geometry}
 \usepackage[doublespacing]{setspace}
 \usepackage{tocloft}
 \usepackage[rm, tiny,center, compact]{titlesec}
 \usepackage{indentfirst}
 \usepackage{etoolbox}
\usepackage{tocvsec2}
 \usepackage[titletoc]{appendix}
 \usepackage{appendix}
 \usepackage{tamuconfig}
\usepackage{rotating}
\usepackage{graphicx}
\usepackage{bm}
\usepackage{bbm}
\usepackage{ulem}
\usepackage{color}
\usepackage[latin1]{inputenc}
\usepackage{amsmath}
\usepackage{amsfonts}
\usepackage{amssymb}
\usepackage{graphicx}
\usepackage{caption}
\usepackage{subcaption}
\usepackage{multirow}
\usepackage{array}
\usepackage{float}
\usepackage{blindtext}
\usepackage{subfig}
\usepackage{placeins}
\usepackage[latin1]{inputenc}
\usepackage{amsmath}

\begin{document}

\renewcommand{\tamumanuscripttitle}{Spintronics in half-passivated graphene }
\renewcommand{\tamupapertype}{Dissertation}
\renewcommand{\tamufullname}{Shayan Hemmatiyan }
\renewcommand{\tamudegree}{Shayan Hemmatiyan}
\renewcommand{\tamuchairone}{Jairo Sinova}
\newcommand{\tamuchairtwo}{Artem Abanov}
\renewcommand{\tamumemberone}{Donald G. Naugle}
\newcommand{\tamumembertwo}{Tahir Cagin}
\renewcommand{\tamudepthead}{Peter McIntyre}
\renewcommand{\tamugradmonth}{December}
\renewcommand{\tamugradyear}{2016}
\renewcommand{\tamudepartment}{Physics}

%
%
%


\providecommand{\tabularnewline}{\\}

\begin{titlepage}
\begin{center}
\MakeUppercase{\tamumanuscripttitle}
\vspace{4em}

A \tamupapertype

by

\MakeUppercase{\tamufullname}

\vspace{4em}

\begin{singlespace}

Submitted to the Office of Graduate and Professional Studies of \\
Texas A\&M University \\

in partial fulfillment of the requirements for the degree of \\
\end{singlespace}

\MakeUppercase{\tamudegree}
\par\end{center}
\vspace{2em}
\begin{singlespace}
\begin{tabular}{ll}
 & \tabularnewline
& \cr
Chair of Committee, & \tamuchairone \tabularnewline 
Co-Chair of Committee, & \tamuchairtwo \tabularnewline 
Committee Members, & \tamumemberone\tabularnewline
 & \tamumembertwo\tabularnewline

Head of Department, & \tamudepthead\tabularnewline

\end{tabular}
\end{singlespace}
\vspace{3em}

\begin{center}
\tamugradmonth \hspace{2pt} \tamugradyear

\vspace{3em}

Major Subject: \tamudepartment \par
\vspace{3em}
Copyright \tamugradyear \hspace{.5em}\tamufullname 
\par\end{center}
\end{titlepage}
\pagebreak{}

%
%
%

\chapter*{ABSTRACT}
\addcontentsline{toc}{chapter}{ABSTRACT} 

\pagestyle{plain} 
\pagenumbering{roman} 
\setcounter{page}{2}

\indent In this thesis, I propose a practical way to stabilize half passivated graphene (graphone). I show that the dipole moments induced by a hexagonal-boron nitride (h-BN) substrate on graphene stabilize the hydrogen atoms on one sublattice of the graphene layer and suppress the migration of the adsorbed hydrogen atoms. I also present the substrate effect of h-BN that reduces distortion induced by fluorination of graphene and stabilizes half-passivated graphene in a single sublattice. Then using spin-polarized density functional calculations I investigate magnetic properties of graphone. I show the system has different magnetic order which can be described by either super-exchange or double exchange mechanisms depending on the type the of add atom. The hybridization of graphene changes from $sp^2$- to $sp^3$- type hybridization due to the buckling induced by passivation. The change in hybridization together with add-atom orbitals induces a fairly large spin orbit coupling (SOC). 
%
Based upon first principle spin-polarized density of states calculations, I show that the graphone obtained in different graphene/h-BN heterostructures exhibits a half metallic state. I propose to use this exotic material for spin valve systems and other spintronics devices. 

\pagebreak{}

%
%
%

\chapter*{DEDICATION}
\addcontentsline{toc}{chapter}{DEDICATION}  

\indent  I dedicate my dissertation work first to my parents for their endless support as a special feeling of obligation to express my gratitude to my loving parents and my greatest teachers Mohammadali and Fatemeh.  They have been actively supporting me to realize my potential and teaching me to never give up on my dreams. 

I dedicate this thesis to my sister Azadeh and my brother Shahrooz for all of their continued love and support. I also dedicate this thesis to all my friends and give special thanks to my best friend Shima for being there for me throughout the doctorate program.

Finally, I dedicate this thesis to any who appreciate the beauty of nature expressed in physical laws.

\pagebreak{}

%
%
%

\chapter*{ACKNOWLEDGEMENTS}
\addcontentsline{toc}{chapter}{ACKNOWLEDGEMENTS}  

\indent Firstly, I would like to express my sincere gratitude to my advisor Prof. Jairo Sinova for the continuous support of my Ph.D study and related research, for his patience, motivation, and immense knowledge. Jairo gave me the freedom and independency to pursue various projects. I could not have imagined having a better advisor and mentor for my Ph.D study.

I am also very grateful to Prof. Artem Ababnov for his scientific advice and knowledge and many insightful discussions and suggestions. His knowledge in condensed matter physics 

Besides my advisor, I would like to thank first my co-advisor Prof. Artem Abanov the rest of my thesis committee:  Prof. Donald G. Naugle, Prof. Tahir Cagin, and Dr. Xiaofeng Qian for their insightful comments and encouragement, but also for the questions which incentivised me to widen my research from various perspectives.

My sincere thanks also goes to Prof. Allan MacDonald and Dr. Marco Polini, who provided me an opportunity to collaborate with them in this project. 

I would like to thank my M.Sc. research advisors, Prof. Abdullah Langari for their constant enthusiasm and encouragement.

Many thanks to my great friends Shima Shayanfar, Nader Ghassemi and Yusef Maleki and my colleagues and friends Dr. Erik McNellis, Dr. Amaury Souza, Dr. Yuta Yamane, Jacob Gayles and Cristian Cernov.  

Last but not the least, I would like to thank my amazing family: my parents Mohammadali and Fatemeh and to my brother Shahrooz and my sister Azadeh for supporting me spiritually throughout writing this thesis and my life in general.

\pagebreak{}
%
%
%
%

\chapter*{CONTRIBUTORS AND FUNDING SOURCES}
\addcontentsline{toc}{chapter}{CONTRIBUTORS AND FUNDING SOURCES}  

\subsection*{Contribusions}
This work was supported by a thesis committee consisting of Professors Jairo Sinova, Artem Abanov and Donald G. Naugle of the Department of Physics and Astronomy and Professor Tahir Cagin of the Department of Material Science and Engineering.

I would like to acknowledge the contribution of my collaborators Prof. Allan MacDonald and Dr. Marco Polini for their insightful comments and discussions. 

I have completed a large portion of this work independently and the results were published in Refs. \cite{PhysRevB.90.035433, hemmatiyan2013vertical, hemmatiyan2014stable, hemmatiyan2015lateral, pan2013vertical, cernov2014screening}.

\subsection*{Funding Sources}
This work was supported by SWAN, DMR-1105512, ONR- n000141110780, Alexander von Humboldt Foundation and  the ERC Synergy Grant SC2 (No. 610115).

\include{nomenclature}

%
%
%

\phantomsection
\addcontentsline{toc}{chapter}{TABLE OF CONTENTS}  

\begin{singlespace}
\renewcommand\contentsname{\normalfont} {\centerline{TABLE OF CONTENTS}}


\setlength{\cftaftertoctitleskip}{1em}
\renewcommand{\cftaftertoctitle}{%
\hfill{\normalfont {Page}\par}}

\tableofcontents

\end{singlespace}

\pagebreak{}


\phantomsection
\addcontentsline{toc}{chapter}{LIST OF FIGURES}  

\renewcommand{\cftloftitlefont}{\center\normalfont\MakeUppercase}

\setlength{\cftbeforeloftitleskip}{-12pt} 
\renewcommand{\cftafterloftitleskip}{12pt}

\renewcommand{\cftafterloftitle}{%
\\[4em]\mbox{}\hspace{2pt}FIGURE\hfill{\normalfont Page}\vskip\baselineskip}

\begingroup

\begin{center}
\begin{singlespace}
\setlength{\cftbeforechapskip}{0.4cm}
\setlength{\cftbeforesecskip}{0.30cm}
\setlength{\cftbeforesubsecskip}{0.30cm}
\setlength{\cftbeforefigskip}{0.4cm}
\setlength{\cftbeforetabskip}{0.4cm} 

\listoffigures

\end{singlespace}
\end{center}

\pagebreak{}

%
\phantomsection
\addcontentsline{toc}{chapter}{LIST OF TABLES}  

\renewcommand{\cftlottitlefont}{\center\normalfont\MakeUppercase}

\setlength{\cftbeforelottitleskip}{-12pt} 

\renewcommand{\cftafterlottitleskip}{12pt}

\renewcommand{\cftafterlottitle}{%
\\[4em]\mbox{}\hspace{4pt}TABLE\hfill{\normalfont Page}\vskip\baselineskip}

\begin{center}
\begin{singlespace}

\setlength{\cftbeforechapskip}{0.4cm}
\setlength{\cftbeforesecskip}{0.30cm}
\setlength{\cftbeforesubsecskip}{0.30cm}
\setlength{\cftbeforefigskip}{0.4cm}
\setlength{\cftbeforetabskip}{0.4cm}

\listoftables 

\end{singlespace}
\end{center}
\endgroup
\pagebreak{}  
%
%
%


\pagestyle{plain} 
\pagenumbering{arabic} 
\setcounter{page}{1}

\chapter[INTRODUCTION]{\uppercase {INTRODUCTION} \footnote[1]{Part of the data reported in this chapter is reprinted with permission from my work in Ref. \cite{PhysRevB.90.035433}} }\label{Introduction}
\section{Outline of This Thesis}
The low electronic density of states near the Fermi-energy and the atomic thickness of graphene make it a very attractive material for high frequency large-scale integrated electronics \cite{geim2007rise}. However, due to its semi-metallic nature (zero band-gap at neutrality point), graphene exhibits a small ON/OFF switching ratio ($<$10 at room temperature). This problem inhibits the application of graphene for charge based logic devices and integrated circuits. \cite{britnell2012field, hemmatiyan2013vertical, pan2013vertical} 

One possibility to overcome the small ON/OFF switch ratio is to control charge current via spin. The proposed mechanisms to adjust the current include applying an external magnetic field or a magnetization switching (charge based) via spin polarized current in spin valve systems and tunneling magneto resistance (TMR) devices.\cite{kiselev2003microwave, bertotti2005magnetization, PhysRevLett.100.186805, PhysRevLett.74.3273}

The magnetic properties of graphene have also been extensively studied
in the recent years.\cite{sorella2007semi, PhysRevB.75.125408, yazyev2010emergence, PhysRevB.77.195428, PhysRevB.78.235435, PhysRevB.77.134114, PhysRevB.84.125410} 
It has been shown, that pure graphene exhibits only weak antiferromagnetic (AF) order \cite{sorella2007semi} at near room temperature. However, studies of structural and other defects show that induced  $sp^3$-type hybridization, such as mono-vacancies create a local spin moment and, in certain circumstances, promote a robust long range magnetic order \cite{PhysRevB.75.125408,yazyev2010emergence}.

A similar effect is obtained by chemical functionalization of the graphene with elements such as hydrogen \cite{PhysRevB.75.125408, yazyev2010emergence, zhou2009ferromagnetism, moaied2014hydrogenation}. 
Maximum magnetic moment (1$\mu_B$ per cell) is predicted for the half hydrogenated-graphene (graphone). In such theoretical systems the hydrogen atoms are adsorbed by only one graphene sublattice. Induced $sp^{3}$ hybridization then results in localized magnetic moment in the same way as for defects. However, the overlap between the $p_{z}$ orbitals of the nearest carbons  is sufficient to create  a long range magnetic order at room temperature \cite{yazyev2010emergence}. The $p_z$ orbitals have a near $3$ eV band-gap \cite{PhysRevB.82.153404, kharche2011quasiparticle} in graphone.

Although graphone could make a breakthrough in spintronics, its fabrication is an experimentally challenging task. Fundamentally, there are two obstacles:
i) the symmetry between sublattices, and ii) the lack of barrier for the trapped hydrogen atoms to migrate between the sublattices. \cite{boukhvalov2010stable}

Recently, there has been some progress in fabrication of partially hydrogenated graphene \cite{peng2014new, giesbers2013interface} under certain experimental conditions (i.e low temperature) \cite{peng2014new}. Zhou et al proposed a functionalized heterostructures of fully hydrogenated graphene (graphane) on the top of hexagonal-boron nitride (h-BN) \cite{zhou2012fabricate}. The idea is to create an active nitrogen agent by exposing the system to fluorine. The instability of nitrogen-fluorine bond increases the electronegativity of nitrogen thus creating an active nitrogen site. By applying pressure on the fluorinated h-BN layer, the system undergoes a structural transition from graphane to semi-hydrogenated graphene by adsorption of all the hydrogen atoms from one sublattice in graphane to the h-BN layer \cite{zhou2012fabricate}. 

Hexagonal boron nitride is the insulating isomorph of graphite (honeycomb lattice with boron and nitrogen on two adjacent sublattices) with a large band-gap of 5.97 eV. It was shown \cite{dean2010boron} that it is a superior substrate for graphene for homogenous and high quality graphene fabrication \cite{dean2010boron, kim2013synthesis, yang2013epitaxial}. The small lattice mismatching ($1.7$\%) and atomically planer structure of h-BN (free of dangling bonds and charge traps) preserve the properties of graphene such as charge carrier mobility.

In this thesis, I present a practical method to solve the two obstacles (symmetry of two sublattices and mobility of hydrogen) using the functionalized graphene hybrid structure with h-BN. Our proposed experimental set up included two steps: fabrication of graphene on h-BN substrate then exposing the system to a hydrogen plasma. The electrical dipole induced by the substrate in addition to small buckling of carbon bonds will trap hydrogen in one sublattice and will kinematically stabilize the system.

I show that for h-BN the difference in electronegativity of nitrogen and boron creates a dipole moment for each nitrogen site. This dipole moment breaks the equivalency of two carbon atoms in two different graphene sublattices. The similar screening effect has been reported in multilayer graphene but with different strength \cite{PhysRevB.91.155419}. Moreover, the screening effect of h-BN will generate a buckling in the graphene layer. This buckling will change the vertical position of the one sublattice with respect to the other sublattice and will enhance the coverage rate of the hydrogen in one sublattice. The dipole moment is also responsible for the increased migration barrier in the adsorbed hydrogen atoms, effectively pinning the hydrogen atoms to one sublattice.
 
I also show that the very same dipole moment induced by the h-BN substrate changes the Fermi-energy of the graphone layer and decreases the band-gap of the graphone from near 3 eV in pristine graphone to 1.93 eV in graphone/h-BN heterostructure \cite{kharche2011quasiparticle}.

\begin{figure}[!h]
\centering
\includegraphics[width=0.8\columnwidth]{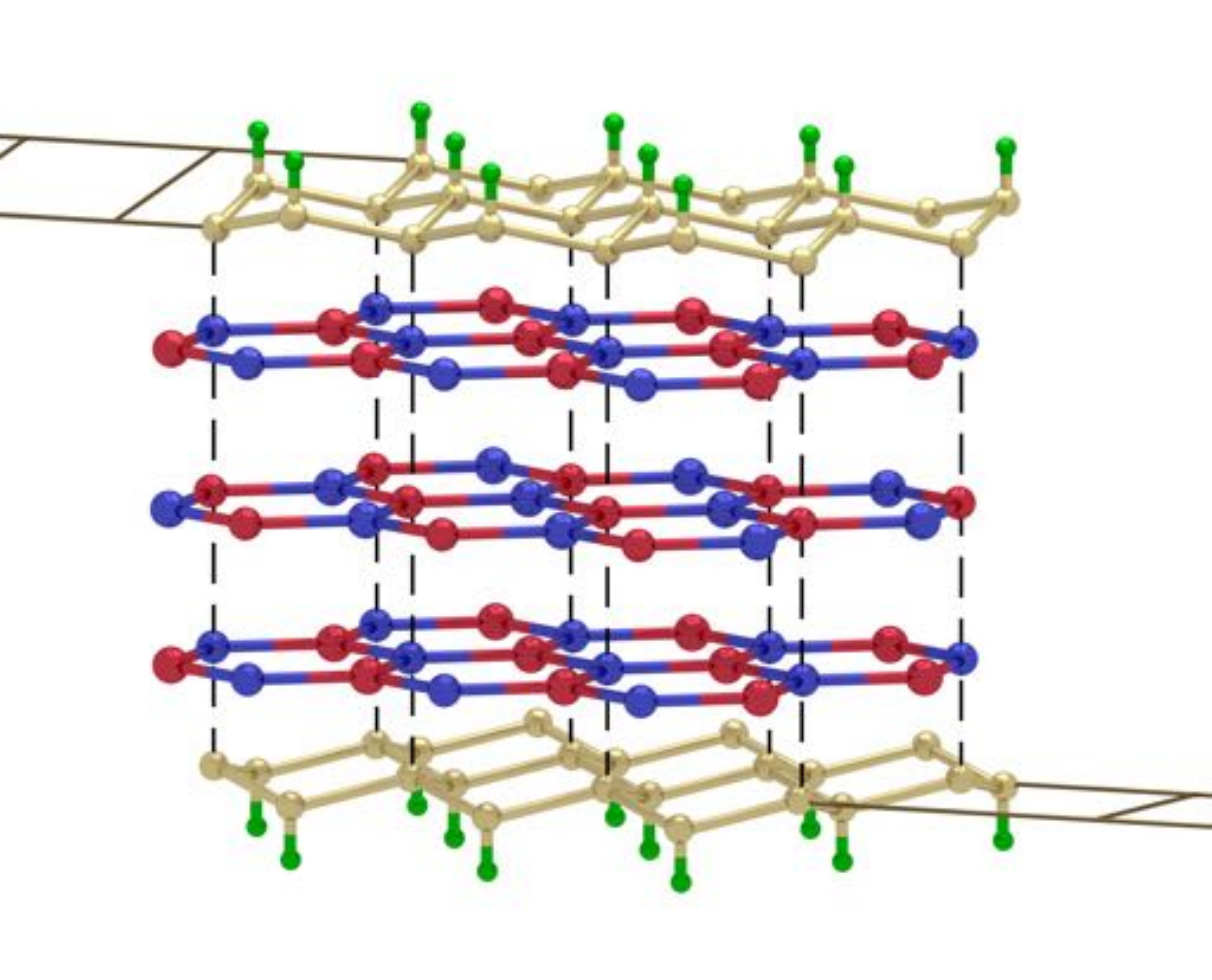}
\caption{(Color online) TMR transistor: Graphone(B)/h- BN/graphone(B), two ferromagnetic layers of graphone, and few (three) layers of h-BN between.} 
\label{fig:tmr3bn}
\end{figure}

I point out that the spin-orbit coupling (SOC) in graphone increases due to a change in hybridization. In graphene, carbon has small SOC due to its $sp^2$-type hybridization. The $d$-orbital has the main contribution in creating a 24 $\mu eV$ intrinsic band-gap \cite{PhysRevB.82.245412}. However, it has been seen that the hydrogen atom enhances spin orbit interactions in graphene via transition from $sp^2$- to $sp^3$- type hybridization due to bending of bonds between carbon atoms  \cite{zhou2010enhanced, PhysRevLett.110.246602}. I show that this effect increases in the presence of the h-BN.

Finally, I propose to use the graphone/h-BN heterostructures in TMR devices as shown in Fig. \ref{fig:tmr3bn}. 
The tunneling current is tuned by changing the magnetization of the one of the electrodes \cite{PhysRevLett.74.3273, kiselev2003microwave, bertotti2005magnetization, PhysRevLett.100.186805}.
I realize that multilayer graphone/h-BN heterostructure is a half metal with near 100\% spin polarization. It was speculated before \cite{julliere1975tunneling} that such materials are ideal for TMR devices.

\section{Thesis Organization}
This thesis is organized as follows: In chapter \ref{chap:dft}, I review basics of density functional theory.  In chapter \ref{chap:graphene}, I give an overview on the fundamental properties of graphene. Then I discuss the magnetic properties of pristine graphene and possible methods to induce magnetization in graphene. 

In chapter \ref{chap:graphone}, I show how the screening effect of h-BN depends on stacking. I calculate the mobility of hydrogen atoms on both pristine graphene and graphene/h-BN heterostructures and I show that the graphone/h-BN heterostructure is stable. I also illustrate that the h-BN reduces distortion effect caused by fluorination in graphene and stabilized half-fluorinated graphene.

 In chapter \ref{chap:magnetization}, I perform spin polarized calculations to determine the stable magnetic state. Then I calculate the magnitude of the SOC (both intrinsic and Rashba) in the presence of h-BN. In chapter \ref{chap:TMR}, I show that not only could the h-BN heterostructure be used to trap hydrogen for fabrication of graphone, but it could also be utilized as an insulator in TMR devices.

%
%
%


%

\chapter{\uppercase {DENSITY FUNCTIONAL THEORY}}\label{chap:dft}
Density functional theory (DFT) is one of the most popular and successful quantum mechanical approaches to matter. It offers a powerful technique to understand the ground state properties of interacting electrons. For example, DFT is applied to calculate band structure of solids, biding energy of the molecules and magnetic properties of the materials. Over the past few decades, DFT was successful in explaining magnetic properties of transition metals and their alloys  \cite{singh2013electronic}. In this chapter we review some aspects of DFT framework.

 DFT method unlike traditional many-electron based electronic structure methods like Hartree-Fock theory uses the electronic density $n(r)$ as the basic parameter. The density $n(r)$ is only a function of three variables. Thus using electronic density instead of many-body wavefunction offers a big reduction in parameter space; whereas the many-body wavefunction has 3N-dependent variables (without including spin degrees of freedom). In other words, DFT reduces the ground state solution of interacting particles into the solution of single-particle Hartree-type equations. 
 
 The goal of this chapter to give an overview of DFT in less detail, but more extensive reviews on DFT have been presented in Refs. \cite{parr1994density, leenaerts2009hydrogenation, kohanoff2006electronic, bickelhaupt2007kohn}. This chapter is organized as follows: in section \ref{sec:review_dft}, basic concepts of DFT are reviewed then in section \ref{sec:NEB}, a minimum energy path method for finding the activation energy and activation barrier is introduced. Finally, spin polarized DFT calculations are discussed in section \ref{sec:SDFT}.

\section{Review on Density Functional Theory (DFT)}\label{sec:review_dft}
To get some idea about DFT, it is useful to recall some elementary quantum mechanics. In quantum mechanics all the information about the system is stored into the system's wavefunction $\Psi$. This wavefunction is the solution to Schr{\"o}dinger equation (SE), which for a single electron moving in a potential $v(r)$ is

\begin{eqnarray}
\big[-\frac{\hbar^2 \nabla^2}{2m} + v(r) \big] \Psi(r) = E \Psi (r) .
\end{eqnarray}

In general we can extend this Hamiltonian to a many-body system. Therefore, many-body Schr{\"o}dinger (non-relativistic) becomes
\begin{eqnarray}
&& H  = T_N +V_{NN}+T_e++T_{e}+ V_{ee}+V_{eN}++V_{ext}  \nonumber \\
&& = \sum_{I} -\frac{\hbar^2}{2M_I} \nabla^{2}_{I} +\sum_{I < L} \frac{N_I N_L e^2}{\vert R_{I} - R_{L} \vert } \nonumber \\
&& +\sum_{i} -\frac{\hbar^2}{2m} \nabla^{2}_{i} +\sum_{i < j} \frac{e^2}{\vert r_{i} - r_{j} \vert } \nonumber \\
&& - \sum_{i, L} \frac{N_L e^2}{\vert r_{i} - R_{L} \vert} + V_{ext},
\end{eqnarray}
where the first term is the  nuclei kinetic energy and the second is the coulomb interactions. Third term and forth terms are, respectively, for the electron kinetic energy and coulomb interaction and the fifth term is for electron-nuclei interaction. Finally, the last term is for an external potential in the system. In quantum mechanics the usual path is as follows:
\begin{eqnarray}
V_{ext} \xrightarrow{SE} \psi(r_1, r_2, ..., r_N) \xrightarrow{\langle \Psi \vert ... \vert \Psi \rangle} observables.
\end{eqnarray}

The many-body Hamiltonian per se contains both electron and nuclei degrees of freedom. However, the nuclei mass is much larger than the  electron mass ($10^3 -10^5$). Thus, the nuclear kinetic energy is relatively small and one can assume the nuclei are frozen (i.e. $T_N = 0$). This can be interpreted as a static effective potential caused by nuclei. This approximation in which electron and nuclei degrees of freedom are separated is called Born-Oppenheimer (BO) approximation. The nuclei position only enters as a parameter in the Hamiltonian from the BO approximation $H(\{R_I\})$. Thus the many-body Hamiltonian is simplified to the Schr{\"o}dinger equation for N-electrons in a static nuclear field applied by nuclei. 

The Schr{\"o}dinger equation for N-electrons with non-local interactions is still complicated. DFT supplies a prescription to deal with the universal operators $\hat{T}$ and $\hat{U}$. It offers a systematic method to map a many-body system with non-local electron-electron interaction ($V_{ee}$) into a single particle moving in an effective field. Two core conceptual theorems called the Hohenberg and Kohn Theorems arise to explain this mapping.

\subsection{Hohenberg and Kohn Theorems}
Too many degrees of freedom in many-body interacting system hinders a generic solution to the Schr{\"o}dinger equation. The electronic many-body wavefunction \\$\Psi(x_1, x_2, ... , x_N)$ is a complex function of 3N-parameters (not including spin degrees of freedom). Moreover, due to the fermionic nature of electrons the many-body wavefunction should be antisymmetric. The electron density of the ground state is denoted as:

\begin{eqnarray}
n(r) = \langle \psi \vert \sum_{i=1}^{N} \delta(r-r_i) \vert \psi\rangle .
\end{eqnarray}

Note that electron density only depends on three position parameters (not 3N). Hohenberg-Kohn shows \cite{PhysRev.136.B864} that 
\begin{itemize}
\item for any interacting particles in an external potential $V_{ext}(r)$, the density is a uniquely determined. In other words there is a one to one correspondence between density and external potential,
\item  energy can be written in terms of a universal functional of the density and the exact ground state is the global minimum of this functional. 
\end{itemize}

\begin{eqnarray}
E[n] = min(E[n]) \Longleftrightarrow \frac{\delta E[n]}{\delta n} \vert_{n=n_0} = 0 .
\end{eqnarray}

The first theorem states that we can reduce the number of degrees of freedom drastically for the wavefunction and retain all the necessary information into the density. From the second theorem we can write:

\begin{eqnarray}
E [n]  = F[n]  + \int n(r) V_{ext} dr .
\end{eqnarray}

The first term $F[n]$ is a universal functional representing the sum of kinetic and coulomb energies and the second term includes the external potential of nuclei and other external fields in the system. These two theorems do not provide any practical scheme to calculate ground-state density and the exact form of the functional F[n] still remains unknown. One year after the Hohenberg-Kohn theorem, Kohn and Sham developed a method to obtain the universal functional F[n] and therefore E[n].

\subsection{The Kohn-Sham Equations} 

W. Kohn and L. Sham (KS) have substituted the full interacting-electronic system by a fictitious non-interaction system which minimizes energy E[n] with respect to density n(r). This way, the ground-state density of the auxiliary system of non-interacting particles $n_s$ and the interacting many-body system n(r) (both in the presence of an external potential $V_{ext}$) are matching. In other words, these auxiliary electrons are moving in an effective potential, which is generated by the other particles (other nucleus and electrons). 

The main advantage of this method is working with the single particle wavefunction rather than the interacting many-body wavefunction. The energy can be written as:

\begin{eqnarray}\label{eq:func}
F[n] =T[n] + E_{H}[n] + E_{XC}[n] + E_{ext}[n] ,
\end{eqnarray}
where the first term is the kinetic energy, the second and the third terms are Hartree and exchange-correlation terms, respectively, and the last term stands for external potential.

In general, exchange-correlation functional $E_{XC}[n]$ is not known explicitly and a reasonable approximation should be made. The contributions from equation \ref{eq:func} can be written as follows:

\begin{eqnarray}
E_{ext} = \int V_{ext} n(r) dr ,
\end{eqnarray}

\begin{eqnarray}
T[n] = -\int \sum_{i=1}^{N} \psi^{*}_{i}(r) \frac{\hbar^{2} \nabla^2}{2m} \psi_{i}(r) dr ,
\end{eqnarray} 

\begin{eqnarray}
E_{H} = \frac{e^2}{8\pi \epsilon_0} \int \frac{n(r)n(r')}{\vert r - r' \vert} dr dr' .
\end{eqnarray}

Ground-state of the system can be obtained by minimizing energy under the normalization constraint for the single particle wavefunction. Variation of the energy-functional respect to $\psi^*$ and introducing Lagrange multipliers $\epsilon_{i}$ result in Kohn-Sham equations:

\begin{eqnarray}\label{eq:KS}
(-\frac{\hbar^{2}}{2m}\nabla^2 + V_{eff})\psi_i(r) = \epsilon_i \psi_i(r) ,
\end{eqnarray}

\begin{eqnarray}\label{eq:eff} 
V_{eff} = V_{ext} + V_{H} + V_{XC} = V_{ext} + 2\int \frac{n(r') }{\vert r -r' \vert } + V_{XC}[n(r)] ,
\end{eqnarray}
where

\begin{eqnarray}
V_{XC} = \frac{\delta E_{XC}[n(r)]}{\delta n(r)} .
\end{eqnarray}

Both $V_{H}$ and $V_{XC}$ depend on electron density and they must to be determined before solving the Kohn-Sham equations. This turns the problem into a self-consistent problem. One needs to start from an initial guess for the density function and substitute it into the equation \ref{eq:eff} to define $V_{eff}$. Then once $V_{eff}$ has been determined, it can be plugged into the Kohn-Sham equation (\ref{eq:KS}), and then it can be solved for a new single particle wavefunction. Then the density function obtained from the single particle wavefunction goes to the next self consistent iteration. The solution of the KS equation can be obtained once the procedure has been repeated a sufficient number of times so that the change density is converging and no further changes are happening. Then in principle the resulting density is governed for both the fictitious non-interacting reference electrons and the exact ground state of interacting many-body system. 

\subsection{Exchange and Correlation Functionals} 

To solve the  Kohn-Sham equations one needs to determine the exchange-correlation (XC) energy functional $E_{XC}$ or, equivalently, its functional derivative $V_{XC}[n]$. In general, the exchange-correlation functional is given by	

\begin{eqnarray}
E_{XC} [n]  = \int n(r) \epsilon_{xc}(n(r), \nabla n, ...) dr .
\end{eqnarray}

However, the exact form of the exchange-correlation energy and potentials still  remains unknown. For this purpose, one needs to approximate $V_{XC}[n]$ as a function of electron density $n(r)$. These approximations turn DFT into practice so that they are focal points for the accuracy of DFT calculations. 

The simplest step in the approximation ladder is starting from a homogenous electron gas (HEG). In this approximation the electron density is replaced by an interacting HEG. This approximation is called the Local Density Approximation (LDA), and it is widely used in density functional theory calculations. The exchange-correlation function can be written as:

\begin{eqnarray}
E_{XC}^{LDA}[n(r)] &&= E_{X}^{LDA}[n(r)] +E_{C}^{LDA}[n(r)]  \\ \nonumber
&&=\int  \epsilon_{XC}^{hom}[n(r)] \ n(r) \ dr =\int (\epsilon_{X}^{hom}[n(r)] + \epsilon_{C}^{hom}[n(r)] ) \ n(r) \ dr .
\end{eqnarray}

The exchange energy $E_{X}$ can be obtained exactly for HEG \cite{dreizler1990density,  leenaerts2009hydrogenation, kohanoff2006electronic}.

  \begin{eqnarray}
E_{X}^{hom}[n(r)] = -\frac{3}{4}  (\frac{3}{\pi})^{\frac{1}{3}} \int  n^{\frac{4}{3}}  dr .
 \end{eqnarray}

In contrast to the exchange energy, calculation of the correlation energy requires more complicated approaches, like the Quantum Monte Carlo (QMC) method \cite{PhysRevLett.45.566}. LDA has been applied to many problems, such as calculations of band-structure and total energy in solid-state physics. However, in quantum chemistry LDA is not accurate enough to give a quantitate description of chemical bonds; thus it requires further improvement. 
 
\begin{figure}[h]
\centering
\includegraphics[width=0.8\textwidth]{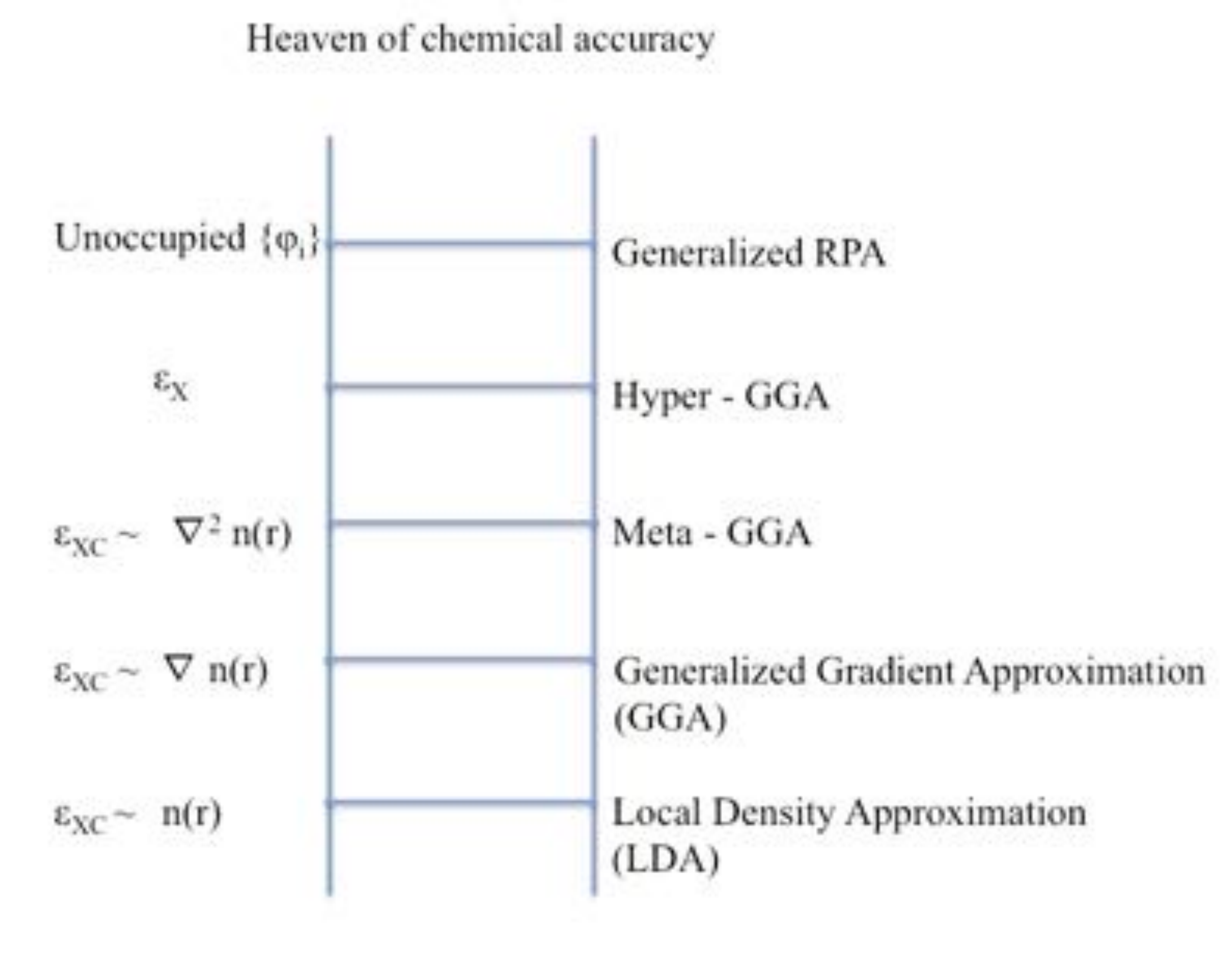}
\caption{Jacob's ladder of the density functional approximation to the $E_{XC}$.} 
\label{fig:JL}
\end{figure}

The next step in the ladder (see figure \ref{fig:JL}) is including the gradient of the density, $\nabla n(r) $, in the expansion of the exchange-correlation potential. This method, the Generalized-Gradient Approximation (GGA) makes the solution of the DFT equations more difficult.

\begin{eqnarray}
E_{XC}^{GGA}[n, \nabla n] = \int F_{XC}[n, \nabla n] \ dr .
\end{eqnarray}

One can also include the density, the gradient of the density and the Laplacian of the density in the expansion of the exchange-correlation energy $E_{XC}$. This type of approximation is called meta-GGA and usually is more accurate than GGA. Hybrid-GGA also increases the accuracy of the results by including part of the exact Hartree-Fock exchange-energy with the GGA exchange-correlational functional. 

\begin{eqnarray}
E^{hyb}_{XC}[n] = a(E_X - E^{GGA}_{X}) +E^{GGA}_{XC} [n] .
\end{eqnarray}

\subsection{The Hellmann-Feynman Theorem}
H. Hellmann in 1937 \cite{hellmann2015hans} and R. Feynman \cite{feynman1939forces} in 1939 presented the Hellmann-Feynman (HF) theorem independently. The theorem states that the expectation value of the electronic Hamiltonian $\hat{H}$ with respect to nuclei positions $\vec{R}_i$ is equivalent to the derivative of the electronic energy E with respect to the nuclei positions $\vec{R}_i$

\begin{eqnarray}\label{eq:deltaE}
\frac{\delta E}{\delta \vec{R}_i} = \frac{\delta}{\delta \vec{R}_i} \langle \psi \vert \hat{H} \vert \psi \rangle = \langle \frac{\delta \psi}{\delta \vec{R}_i} \vert \hat{H} \vert \psi \rangle + \langle \psi \vert \hat{H} \vert \frac{\delta \psi}{\delta \vec{R}_i} \rangle + \langle \psi \vert \frac{\delta \hat{H}}{\delta \vec{R}_i} \vert \psi \rangle ,
\end{eqnarray}
where $\psi$ is the eigenstate of the $\hat{H}$. From normalization condition, we have

\begin{eqnarray}\label{eq:norm}
\frac{\delta }{\delta \vec{R}_i} \langle\psi \vert \psi \rangle = 0 ,
\end{eqnarray}
so that from equation \ref{eq:norm} we can eliminate the first two terms in equation \ref{eq:deltaE} and simplify it to a so called force equation

\begin{eqnarray}
\vec{F}_i \equiv - \frac{ \delta E}{\delta \vec{R}_i} = - \langle \psi \vert \frac{ \delta \hat{H}}{\delta \vec{R}_i} \vert \psi \rangle ,
\end{eqnarray}
where in the above equation $\vec{F}_i$ is the force acting on nuclei $i$. The force theorem plays a crucial role in determining optimized geometry and minimum energy.  The total energy of the system is decreasing by moving in the direction of the force. 

\section{Nudge Elastic Band (NEB) Calculations }\label{sec:NEB}
The Nudge Elastic Band (NEB) method is employed to obtain saddle points and minimum energy path (MEP) between two minima called the reactant and the product. The idea is discretize the initial trajectory (trial) in the configurational space between two minima and relax each of the resulting configurations in the perpendicular directions to the trajectories. Each NEB optimization cycle includes energy and gradient evaluations for a sequence of structures (images). The final NEB gradient is constructed using spring forces that connect the images. 

The relaxation of each point depends on the configuration of the other points (tangent to each point) so that the relaxation towards MEP needs to be performed simultaneously for all points along the path.

\section{Spin-Polarized Density Functional Theory Calculations}\label{sec:SDFT}
The initial DFT formalism was started with non-spin polarized density functional calculations. Without any restriction DFT can also be extended for magnetic systems. Now the ground state of the system depends on both the electron density and the magnetization density $m(r)$. One can write,

\begin{eqnarray}
E = E[n,m] \geq E[n_0, m_0] .
\end{eqnarray}

To include spin, one can replace the wavefunction by a two component spinor,

\begin{eqnarray}
\psi_i(r) = \begin{bmatrix} \psi_{i \uparrow}(r) \\ \psi_{i \downarrow}(r)  \end{bmatrix}
\end{eqnarray}

and extend Kohn-Sham equations by 

\begin{eqnarray}
(-\frac{1}{2}\nabla^2 + V_{eff}(r) +B_{eff}.\sigma ) \psi_{i}(r) = \epsilon_{i} \psi_{i}(r) ,
\end{eqnarray}

\begin{eqnarray}
\rho_{\alpha \beta} (r) = n(r)  \delta_{\alpha \beta } + m(r). \sigma_{\alpha \beta} ,
\end{eqnarray}
where $\sigma \equiv (\sigma_x, \sigma_y, \sigma_z) $ are Pauli matrices. 

\begin{eqnarray}
\sigma_x = \begin{bmatrix} 0 & 1\\1&0\end{bmatrix} , \sigma_y = \begin{bmatrix} 0 & -i\\i&0\end{bmatrix},\sigma_x = \begin{bmatrix} 1 & 0\\0&-1\end{bmatrix},
\end{eqnarray}

\begin{eqnarray}
B_{XC} = \frac{\delta E_{EX}[n,m]}{m(r)} ,
\end{eqnarray}
where
\begin{eqnarray}
n(r) = \sum_{i}^{occ.} \psi^{\dagger}_i \psi_i 
\end{eqnarray}
and
\begin{eqnarray}
m(r) = \sum_{i}^{occ.} \psi_{i}^{\dagger} \sigma \psi_{i} .
\end{eqnarray}

The local Spin Density Functional Approximation is an extension to LDA by including the magnitude of magnetization (local nature of the functional) in addition to the electron density.

\begin{eqnarray}
E_{XC}[n,m] = \int dr \ \epsilon_{LDA}(n(r), \vert m(r) \vert ) . 
\end{eqnarray}

 LDA gives good results for  systems with slowly varying densities. For  systems with stronger changes in electron densities, the spin polarized Generalized Gradient Approximation (GGA), gives a better approximation with the following exchange-correlation energy functional:

\begin{eqnarray}
E_{XC} [n,m]= \int  dr \ \epsilon_{GGA} (n(r), \vert m(r) \vert , \vert \nabla n(r) \vert ) ,
\end{eqnarray}
which does not include gradient in magnetization.
where $\psi_{\alpha}$ and $\psi_{\beta}$ are the corresponding two spin projections.

After this general overview on DFT, the next chapter explains method in the DFT framework more specifically.

\chapter[METHODOLOGY]{\uppercase {METHODOLOGY} \footnote[1]{Part of the data reported in this chapter is reprinted with permission from my work in Ref. \cite{PhysRevB.90.035433}} }
\label{sec:methods}
In this section, I report my methodology exploited in this thesis. All calculations have been done within the first principle framework of the {\sc Quantum ESPRESSO} package \cite{QE-2009} to obtain optimized geometry, Nudged Elastic Band (NEB) and collinear spin polarized DFT calculations. 


\section{Geometry Optimization}
I utilize both the local density  \cite{PhysRevB.23.5048} (LDA) and the generalized gradient approximations \cite{PhysRevLett.77.3865} (GGA) for graphone on top of h-BN. In addition to GGA, van der Waals (VDW) interactions have been treated through VDW-DF2 code \cite{langreth2009density} within the {\sc Quantum ESPRESSO} package. 

It was shown \cite{PhysRevB.77.035427} previously, that LDA is accurate for calculations of the interlayer binding in graphite and multi-layer graphene, while GGA  better matches the experimental data for intra-band interactions and structure. I note, however, that the discrepancy between the two methods was small and was not essential for the main results of this discussion.

LDA calculations were done under ultra-soft, norm-conserving Perdew-Zunger \cite{PhysRevB.23.5048} (PZ) exchange-correlation with the energy cut-off of 60 Ry. For finding the optimized structure and activation energy, I used a large supercell, $5 \times 5$ unit cells, to prevent overlapping between distorted areas. Also, a 20 {\AA} vacuum space was used to avoid interactions between the two periodic layers. 

To find the minimum-energy paths through the migration barrier for the hydrogen atoms, the strong distortion of covalent bonds must be taken into account.
For this purpose, we used nudged-elastic-band (NEB) method outlined in Reference \cite{boukhvalov2010stable} and discussed in section \ref{sec:NEB}.

To find the optimized structure, I used the conjugate-gradient (CG) method with the force and energy convergence parameters $10^{-3}$ Ry/a.u. and $10^{-5}$ Ry, respectively (for both half-hydrogenated and half-fluorinated). A mesh of $4\times4\times1$ k-points in the Mokhorst-Park method \cite{monkhorst1976special} and a cold smearing \cite{PhysRevLett.82.3296} of 0.01 eV degauss (smearing width) was implemented.

\section{Spin-Polarized Calculations}
\subsection{Colinear Magnetization}
For the calculation of the collinear magnetic properties, we used a spin-polarized LDA (LSDA) with a fully relativistic and norm-conserving Perdew-Burke-Ernzerholf (PBE) method \cite{PhysRevLett.77.3865} which has non-linear core corrections; energy and density cut-offs were 50 eV and 500 eV, respectively. I carried out non-collinear calculations using Quantum Espresso in the presence of SOC with all spins constrained along the z-direction (collinear magnetization). My supercell consists of $2\times2$ unit-cells. 

To find the SOC, I fit the band structure of graphone/h-BN with SOC to the tight binding model near the Dirac point. A Marzari-Vanderbilt (MV) smearing of 0.01 eV has been used for calculating density of states. The force and energy optimization accuracy were $10^{-3}$ Ry/a.u. and $10^{-5}$ Ry, respectively. 


\chapter{\uppercase {REVIEW ON GRAPHENE}}\label{chap:graphene}
Graphene with its unique properties has attracted great attention recently. Due to its remarkable electronic, mechanical, optical, thermal and possibly magnetic properties, it is promising for a wide range of nanotechnological applications. In this chapter, some fundamental properties of graphene in relevance to the scope of this thesis are reviewed. Then, I will discuss the effect of defects to introduce magnetization in graphene, followed by the discussion of  half-passivated graphene (graphone).

\section{Fundamental Properties of Graphene}
Carbon with an atomic number of 6 is one of the most (4th) abundant elements in the universe. Carbon with 4 electrons in its valence band bind in different ways, allotropes of carbon. Graphite, diamond and amorphous carbon are some of the most well known allotropes of carbon. In 2004, Andre Geim and Konstantin Novoselov isolated single layer(s) of graphite atoms called graphene using the  scotch tape method and measured their optical and electronic properties. Because of this groundbreaking discovery, Geim and Novoselov won the  2010 Nobel Prize in Physics.

\subsection{Electronic Properties of Graphene}
Graphene due to its unique electronic structure has several potential applications in electronics. As we see in this section, graphene is a gap-less semiconductor (semi-metal) with a linear band-structure with respect to momentum and a zero density of states (DOS) at the Fermi-energy. Due to the zero DOS at the Fermi-energy, the charge carrier conductivity is low.

 The Fermi-energy can be tuned by a gate voltage or by doping (or other type of defects) effectively creates a material with higher conductivity. Graphene is well-known for having a very high charge carrier mobility due to the low density of states at the Fermi-energy (low number of charge carriers). For example, the mobility of the charge carrier for exfoliated graphene on $SiO_2$-covered silicon wafers is 10,000-15,000 $cm^2 V^{-1} s^{-1}$ \cite{chen2008intrinsic}. In principle, the charge carrier mobility can go up to 70,000 $cm^2 V^{-1} s^{-1}$ \cite{chen2008intrinsic, chen2009dielectric}. 

In particular, graphene-based transistors are considered potential candidates to replace silicon electronics in both logic and radio-frequency applications. Graphene-based electronics offer the possibility of making extremely thin field-effect transistors integrated in electronic devices.

However, there is a main challenge which hampers the application of graphene-based transistors in the fabrication of logic-transistors. The main reason is that ideal graphene does not have any band-gap; thus devices and channels based on graphene cannot be switched off \cite{schwierz2010graphene, novoselov2012roadmap}. There are several ways to overcome the switch-off issue of graphene-based transistors. One is opening the band-gap using defects (graphene nano-ribbons \cite{yang2007quasiparticle} or by applying strain \cite{pereira2009tight, ni2008uniaxial} and add-atoms \cite{cervantes2008edge, tang2013tunable}) in graphene. However, the honeycomb structure of graphene is deformed by such methods and results in destruction of the intrinsic properties of graphene (linear band-structure and high charge carrier mobility).

In addition to a defect induced band-gap, Britnell and his coworkers recently have exploited the low density of carriers in graphene (near the Fermi-energy) to build a graphene-based tunneling field effect transistor. Atomically thin boron-nitride or molybdenum-disulfide has been utilized in these graphene heterostructures as a tunneling barrier between two graphene sheets \cite{britnell2012field}. This method preserves the unique properties of graphene and exhibits relatively large ON-OFF switching ratios (50 for boron-nitride and 10,000 for molybdenum-disulfide as a tunneling barrier).

 I also proposed the use of twisted graphene/hexagonal-boron nitride heterostructures to increase ON-OFF switching ratios by one order of magnitude \cite{hemmatiyan2013vertical, pan2013vertical}. In section \ref{chap:TMR}, I propose another strategy to utilize the spin degrees of freedom to control the current in graphene based heterostructures which is the main scope of this thesis.

\subsubsection{Graphene Band-Structure}
\begin{figure}[h]
\centering
\includegraphics[width=0.5\textwidth]{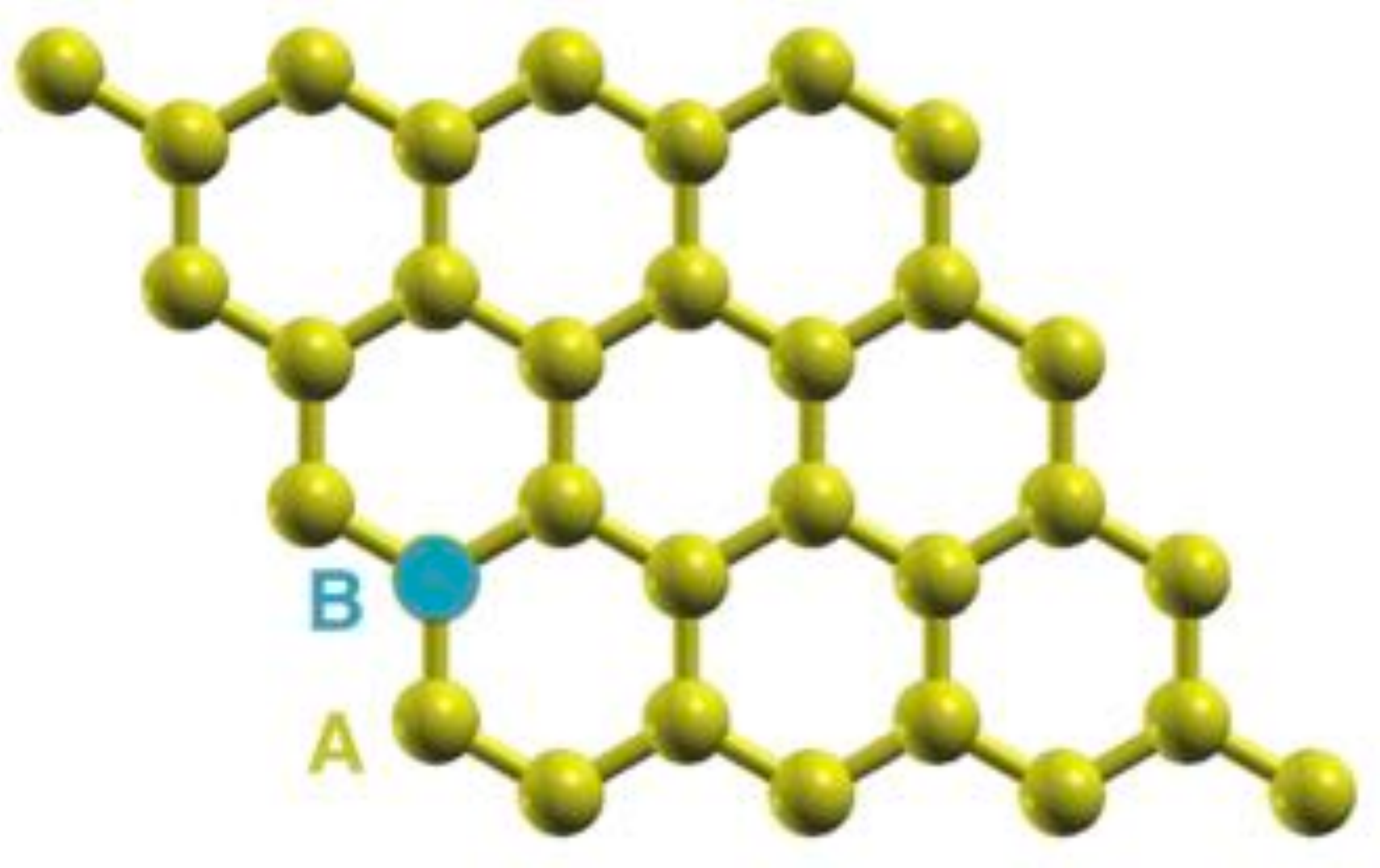}
\caption{Graphene: Honeycomb lattice with two sublattices $A$ and $B$.} 
\label{fig:graphene}
\end{figure}

Graphene is an atomically-thick single layer of graphite with a honeycomb lattice (see figure \ref{fig:graphene}). It has two atoms per unit-cell which form two sublattices (bipartite lattice) as presented in figure \ref{fig:graphene}.

\begin{figure}[h]
\centering
\includegraphics[width=12cm]{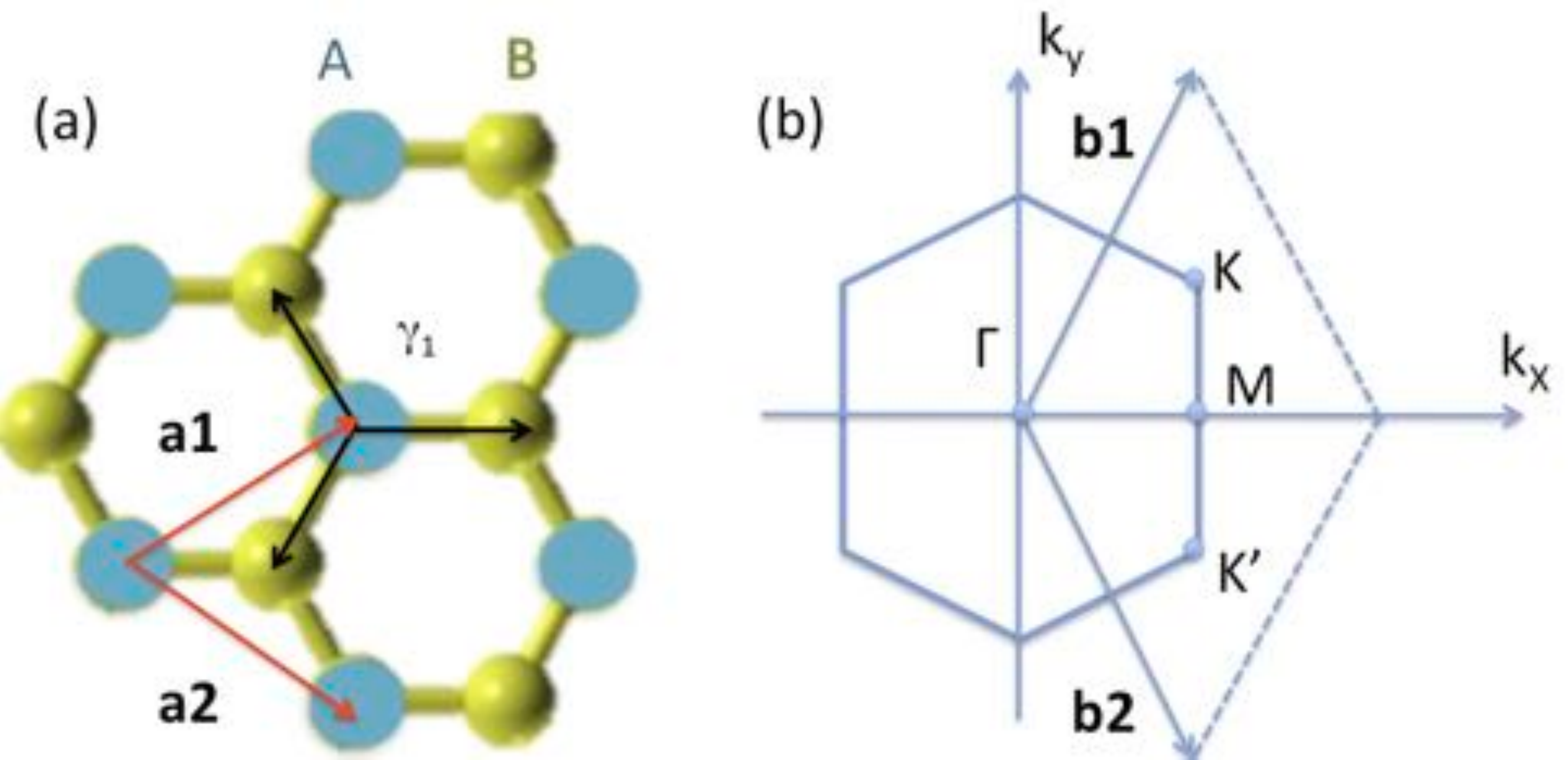}
\caption{Honeycomb lattice structure in graphene (a) and first Brillouin zone (1BZ) of the honeycomb lattice (b). a) The vectors $\textbf{a1}$ and $\textbf{a2}$ are the lattice unit vectors in graphene and $\gamma_1$ is the nearest-neighbor hopping parameter. b) The vectors $\textbf{b1}$ and $\textbf{b2}$ are the reciprocal lattice vectors. The points $\Gamma$, $K(K')$ and $M$ are the center of 1BZ, the Dirac points and the middle of the 1BZ edges, respectively. } 
\label{fig:unit}
\end{figure}

Graphene has a triangular or hexagonal lattice structure with the following unit vectors in the real space:

\begin{eqnarray}
&&\textbf{a1}=\frac{1}{2}( \sqrt{3} \hat{i} +\hat{j} ) \\
&&\textbf{a2}=\frac{1}{2}( \sqrt{3} \hat{i}- \hat{j} ),
\end{eqnarray}
where $a = 2.42 \AA$ is the graphene lattice constant. Followed by unit vectors (see figure \ref{fig:unit}), reciprocal lattice vectors are:

\begin{eqnarray}
&&\textbf{b1}=\frac{2\pi}{\sqrt{3}a}( \hat{i} +  \sqrt{3}  \hat{j}) \\
&&\textbf{b2}=\frac{2\pi}{\sqrt{3}a}(\hat{i} -\sqrt{3}  \hat{j}) .
\end{eqnarray} 

High symmetry points in graphene are the $\Gamma$, $M$ and $K$ points (figure \ref{fig:unit} corresponding to center, center of the edge and corner of first Brillouin zone (BZ), respectively. Carbon in graphene has $sp^2$-type hybridization where each carbon atom has three $\sigma$ bonds with three nearest neighbors. The remaining unpaired electron of each carbon atom forms a half-filled $\pi$-bond with the other unpaired adjacent $p_z$ carbon orbitals. Since these half-filled $\pi$-bands in graphene are delocalized, the system is called $\pi$-conjugated. These $\pi$-bands are responsible for most of the peculiar electronic properties of graphene. In this thesis, some of these properties have been summarized, but, for a more comprehensive review see Refs. \cite{geim2007rise, RevModPhys.83.407, RevModPhys.81.109, balandin2008superior}.

Due to symmetry, the six points at the edge of the BZ  are reduced into two points $K$ and $K'$ in the reduced BZ where the density of states (DOS) vanishes at these points. The electronic band-structure of graphene has been first derived by Wallace \cite{PhysRev.71.622} in 1947 as a starting calculation for band-structure of graphite. Applying the nearest neighbor tight-binding method (in figure \ref{fig:unit} including only $\gamma_1$, called the nearest-neighbor hopping energy) for graphene, one can derive the approximate band-structure of graphene for $\pi$ (bonding) and $\pi^*$ (anti-bonding) bands:

\begin{eqnarray}
E_{\pm}(k_x, k_y) = \pm \gamma_1 (1+4 cos(\frac{3k_xa}{2})cos(\frac{k_ya}{2})+4 cos^2(\frac{k_ya}{2}))^{1/2}.
\end{eqnarray}

Expanding the band-structure of graphene near $K$ ($K'$) points results in a linear dispersion relation (with respect to momentum) which creates the rich physics in graphene and graphene isomorphes (honeycomb lattice based materials):

\begin{eqnarray}
E_{\pm} = \pm \hbar v_{f} \vert k - K \vert,
\end{eqnarray}

\begin{figure}[!h]
\centering
\includegraphics[width=0.8\textwidth]{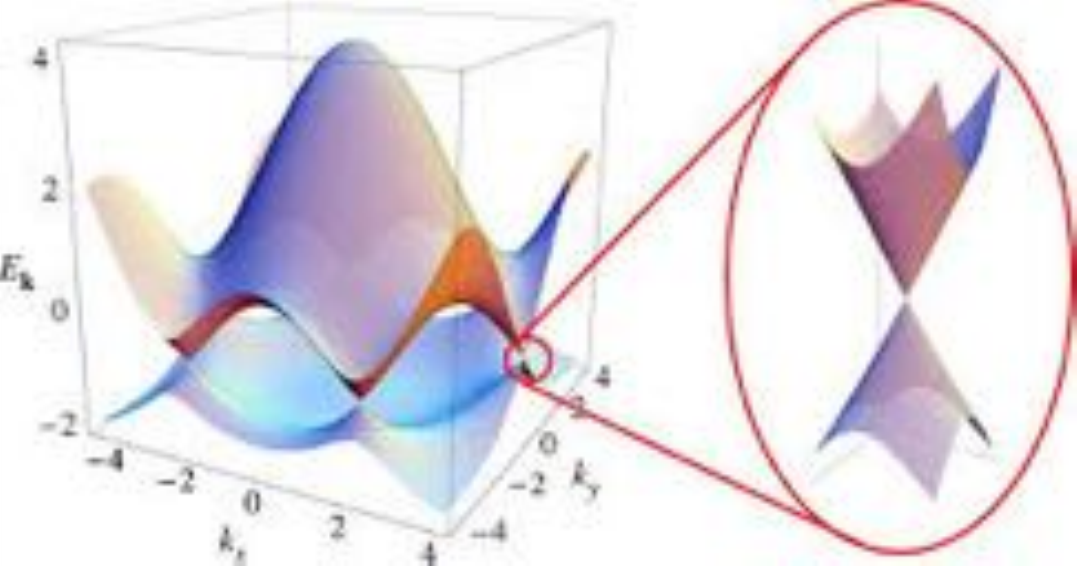}
\caption{Graphene band-structure: Linear dispersion relation near Dirac points lead to 6-identical cones called Dirac-cones.} 
\label{fig:gbslin}
\end{figure}

where $v_f$ is the Fermi-velocity $v_f = 1.1 \times 10^6 \ m/s$. The band-structure of graphene can be represented by two cones. One belongs to the $\pi$-band and the other to $\pi^*$ band. These bands merge at six points of the BZ called the Dirac points or charge neutrality points (see Fig. \ref{fig:gbslin}). At these points (Dirac points) the band-structure, unlike, the parabolic behavior in conventional semiconductors, is in the momentum $k$ (linear dispersion relation). Thus, these points are named Dirac points. This is the main reason why the charge carriers in graphene can be treated as massless Dirac fermions. \cite{novoselov2005two} 

From the energy spectrum, it can be seen that the band-gap of ideal graphene is zero and Fermi-energy resides at Dirac-points (charge neutrality points). This shows that graphene is a semi-metal with a zero band-gap and zero density of states (DOS) at the Fermi-energy (ideal semi-metal). The DOS near the Dirac points can be written:

\begin{eqnarray} \label{DOSg}
\nu(E) = \ \frac{g_s g_\nu \vert E \vert}{2 \pi \hbar^2 v_{f}^{2}},
\end{eqnarray}
where $g_{s}$ and $g_{\nu}$ are the spin and the valley degeneracies, respectively. The electron density $(n)$ and hole density $(p)$ close to the Fermi-energy can be obtained by
\begin{eqnarray}
n(E_f)= \int_{0}^{E_f} \nu(\epsilon) f_{FD}(\epsilon) d\epsilon  \overset{T=0}{\longrightarrow} \frac{g_s g_{\nu}E_{f}^{2}}{4 \pi \hbar^2 v_{f}^2},
\end{eqnarray}

\begin{eqnarray}
p(E_f)= \int_{0}^{E_f} \nu(\epsilon )(1-f_{FD}(\epsilon)) d\epsilon  \overset{T=0}{\longrightarrow} \frac{g_s g_{\nu}E_{f}^{2}}{4 \pi \hbar^2 v_{f}^2},
\end{eqnarray}
where $f_{FD} $ is Fermi-Dirac distribution $f_{FD}(\epsilon) = \frac{1}{1+ e^{\beta (\epsilon - E_{F})}}$ with $\beta = \frac{1}{K_BT}$.

\subsection{Graphene Mechanical Properties}
Existence of a thermodynamically stable 2D material was ambiguous for several years. The reason is that the melting temperature of the materials is decreased by shrinking the thickness. For example, silicon becomes extremely unstable for a thicknesses less than a few atomic layers. The challenge remained unanswered until discovery of the graphene which not only is thermodynamically stable even at the room temperature, but also is the world strongest material. The three $\sigma$ bonds in graphene are the main reason for graphene's robustness. Pure defect-free graphene has a high stiffness (300 - 400 $N/m$) with a breaking strength of $\sim42 N/m$ which is 200 times greater than the steel \cite{cai2009mechanical}.

A large Young's modulus (0.5 - 1.0 $TPa$) in addition to a low weight and remarkable elastic properties, offer a strong and flexible material with several potential applications in aerospace, electronics and medicine. Full review on the potential applications of graphene can be found in the Refs. \cite{choi2010synthesis, allen2009honeycomb}.

\subsection{Graphene Optical Properties}
Single layer of graphene is quite transparent and only absorbs 2.3 \% of the white light. However, considering the thickness in graphene (the atomically thick), such absorption is remarkably high. The transparency together with the high conductivity enables the application of graphene in the fabrication of the conducting transparent electrodes. So graphene is a potential candidate to replace the costly indium thin oxide (ITO) based films. 

\subsection{Graphene Fabrication}
Graphene, first has been discovered an the experiment by K. Novoselov and A. Geim in 2004 \cite{novoselov2004electric} via the exfoliation using a scotch tape. Adhesive tapes helps to peel off the graphite layers. By repeating peeling the graphite over and over, graphite is cleaved into few-layers of graphene. Now there are several more efficient methods to prepare graphene in experiment. Here we only review some of the most important methods.
\begin{itemize}
\item Mechanical exfoliation: this is the most common approach to prepare graphene for the laboratory (not suited to electronic industry) \cite{novoselov2004electric}. The graphite layers are bonded together with a week Van der Waals interaction. The exfoliation and cleavage methods use a mechanical and chemical force to break these week bonds and isolate single layer of the graphene from the graphite. \cite{choi2010synthesis}
\item Epitaxial growth: in this method, graphene is epitaxially grown on the metals. For example, the Chemical Vapor Deposition (CVD) of graphene on Ni or on Cu. \cite{bae2010roll} This method is suitable for the fabrication of a relatively large graphene flakes ($\sim cm^2$) \cite{reina2008large}.
\item Thermal decomposition of SiC: this method requires an ultra-high vacuum (UHV) and single crystal substrate \cite{PhysRevB.78.245403} so it is a costly method. 
\end{itemize}



\section{Direct Evidence of Magnetization in Graphene}\label{sec:defect}
Two-dimensional ferromagnets with their fundamentally different magnetic and electronic properties have received a great attention and are highly desirable in spintronics and bio-devices. The magnetic coupling and magnetic moments depends on the coordination number and the dimensionality-related quantum confinements. Therefore, the number of dimensions evidently changes the magnetic properties of the system. 

The graphene magnetic properties have been substantially studied in the recent years \cite{sorella2007semi, PhysRevB.75.125408, yazyev2010emergence, PhysRevB.77.195428, PhysRevB.78.235435, PhysRevB.77.134114, PhysRevB.84.125410}. According to the Mermin-Wagner theorem \cite{mermin1966absence}, ferromagnetic order cannot exist at any finite temperature in one- and two-dimensional isotropic systems. Therefore, ideal graphene (2D isotropic system) per se does not maintain any trivial magnetic order.

The pure graphene exhibits only a weak antiferromagnetic (AF) order at near room temperature \cite{sorella2007semi}. However, the studies show the magnetic order can exist in graphene by breaking its symmetry and introducing intrinsic anisotropy \cite{PhysRevB.75.125408,yazyev2010emergence}. For example, some defects create an imbalance between two graphene sublattices and breaks the sublattice symmetry. 

There are two types of the defects in graphene, vacancy and chemisorbed atoms. The vacancies, like single vacancy and edge/boundaries create a non-zero spin-density (near the vacancies) \cite{berashevich2010sustained, paz2013connection, yazyev2008magnetism, kumazaki2008local, kan2010ferrimagnetism, palacios2008vacancy}. 

\begin{figure}[h] 
\centering
\includegraphics[width=0.7\textwidth]{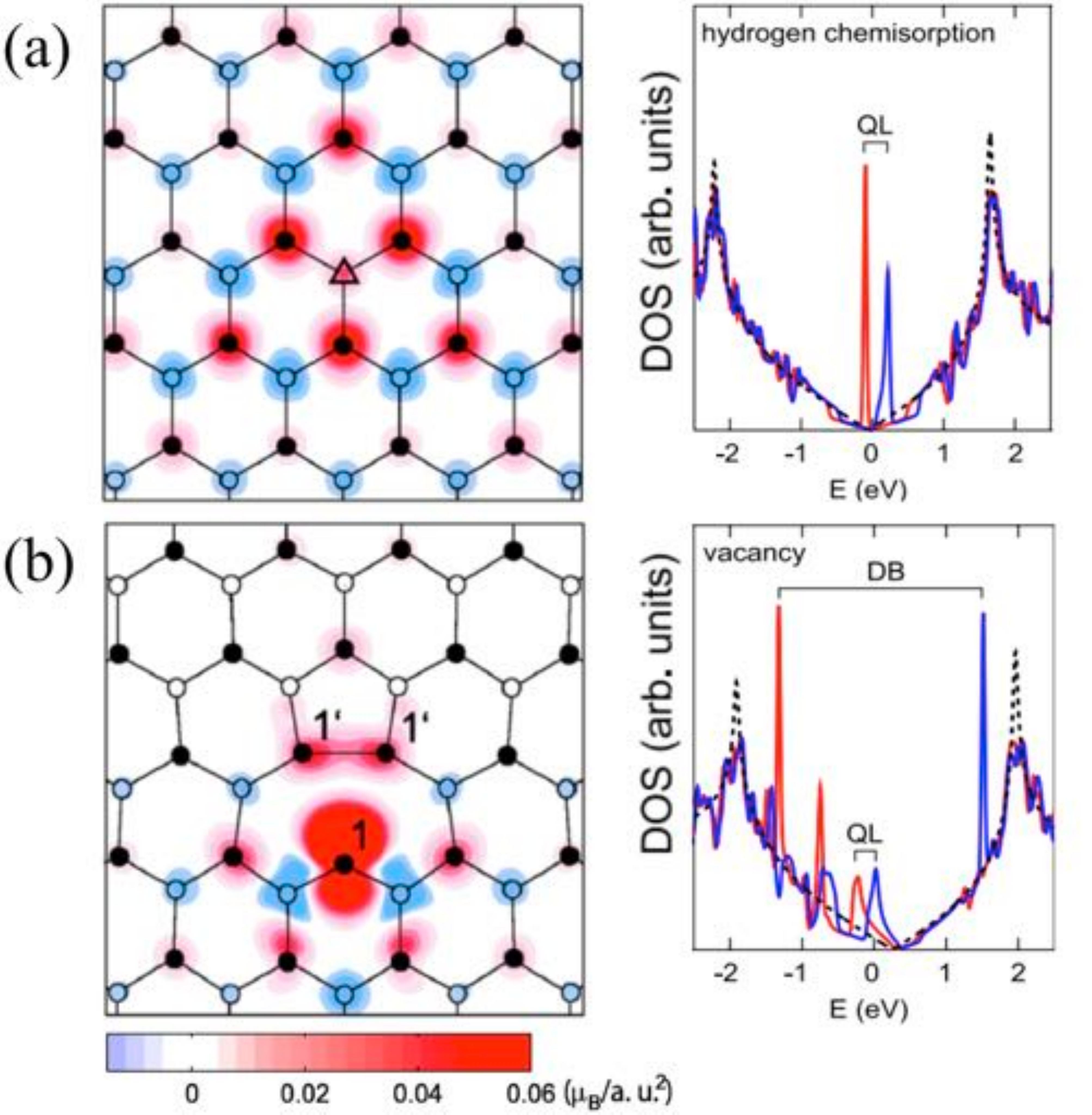}
\caption{Defects induced magnetic moments: The spin density projection (in $\mu_B/a.u.^2$ from around the (a) hydrogen chemisorption defect ($\Delta$) and (b) vacancy \cite{yazyev2007defect} (left figure).The spin-resolved DOS first principle calculation \cite{yazyev2010emergence} (right figures) for the (a) single hydrogen passivation (b) vacancy; dashed curve shows the reference graphene DOC, the red curve is for the majority and the blue curve for the minority. In the right figures, the splitting between the quasi-localized (QL) state peaks (for hydrogen adsorption) and dangling-bond (DB) states (for vacancy) shows the exchange splitting \cite{yazyev2010emergence}. Figures are reprinted with permission from Ref. \cite{yazyev2010emergence} and Ref. \cite{yazyev2007defect} }
\label{fig:def}
\end{figure}

Apart from vacancies, it has been shown that the non-magnetic chemisorbed atoms and complexes such as H, F, O, -CN, -OH \cite{yazyev2010emergence, kumazaki2007nonmagnetic, yazyev2007defect,uchoa2008localized, neto2009adatoms} create a spin imbalance; thus, resulting in the spin-dependent quasi-localized states close to the Fermi-level (see figure \ref{fig:def}).

In both single vacancy and chemisorption defects, the magnetization is induced due to the presence of the quasi-localized defect states \cite{PhysRevB.75.125408}. In the case of hydrogen chemisorption, it gives rise to the strong Stoner ferromagnetism (see figure \ref{fig:def}) with a $1\mu_B$ magnetization per defect \cite{aharoni1996introduction}. The defect band leads to a narrow peak which stabilizes the magnetic order at a high currie temperature \cite{PhysRevB.75.125408}. 

However, the vacancy-type defects show relatively small stoner parameters (small energy splitting between the majority and the minority defect peak states) as it is shown in the figure \ref{fig:def}. It has been illustrated the vacancy-type defects result in a weak stoner ferromagnet with a fractional magnetization (1.15 $\mu_B$ per vacancy) \cite{PhysRevB.75.125408}.

\begin{figure}[h]
\centering
\includegraphics[width=0.7\textwidth]{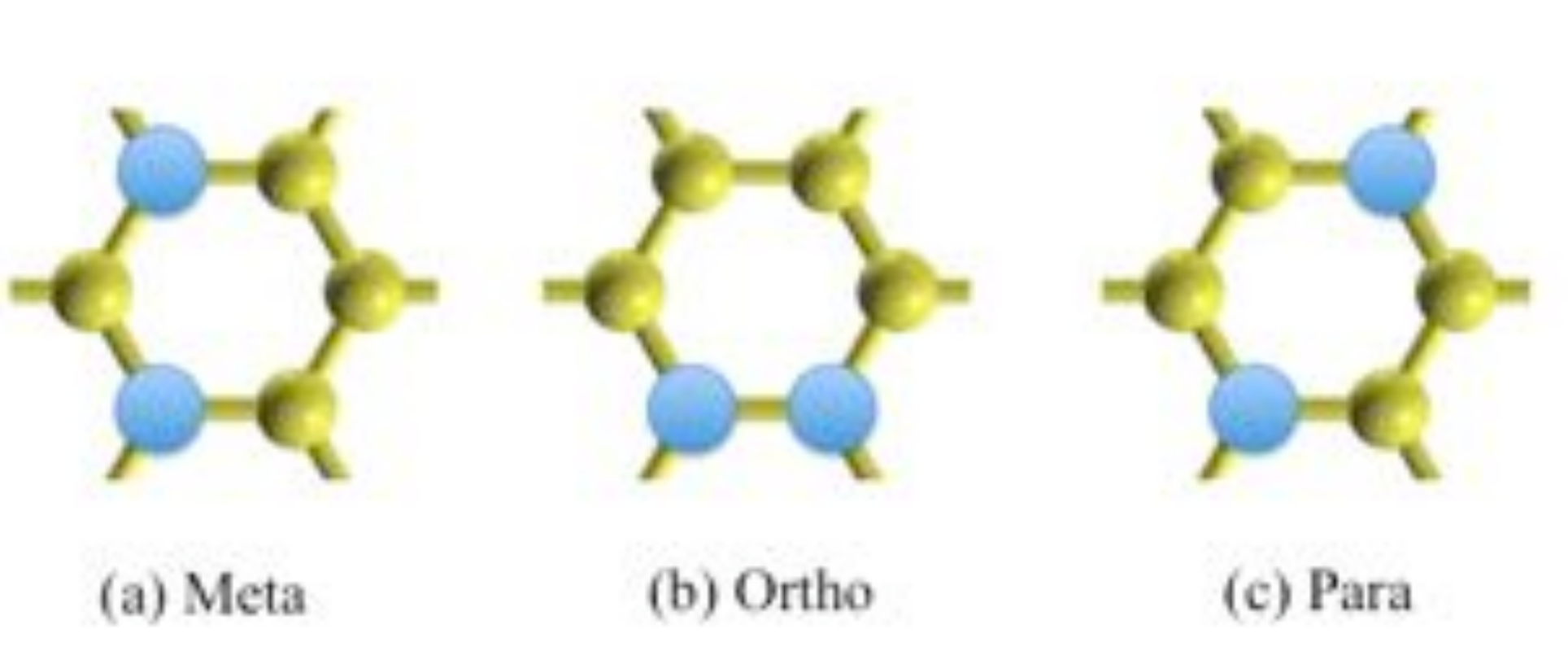}
\caption{Different type of dimer position (two adatoms) on graphene: (a) Metha, (b) ortho and (c) para} 
\label{fig:meta}
\end{figure}

The net induced magnetic moment due to defects in general depends on the density, type and relative position of defects \cite{yazyev2010emergence}. For example, the  ferromagnetic order occurs when two hydrogen atoms are adsorbed to the same sublattice in graphene (meta configuration in figure \ref{fig:meta}). However, the net magnetization is zero for the case of ortho and para \cite{yazyev2010emergence}. Therefore, the magnetization increases once the hydrogen atoms cover more carbon atoms belonging to the same sublattice. In general, the average magnetization $\bar{M}$ can be written as $\bar{M} = \mu_B \vert   N_A^d - N_B^d \vert$ where $N_A^d$ and $N_B^d$ are the number of defects in sublattice A and B, respectively \cite{yazyev2010emergence}.

The maximum magnetization ($\mu_B$/unit-cell) can be obtained in the half-passivated graphene in a single sublattice or graphone where all the carbon atoms belong to the same sublattice are covered by adatoms. From the first principle calculations it has been shown \cite{zhou2009ferromagnetism}, half-hydrogenated graphene in a single sublattice is a ferromagnetic semiconductor with near 1 $\mu_B$ per unit-cell. However, D. W. Boukhvalov has shown that due to the practical issues fabrication of the half-hydrogenated graphene in a single sublattice is experimentally challenging \cite{boukhvalov2010stable}. He also shows the magnetic order depends on the type of passivation. For example, the half-fluorinated graphene shows an anti-ferromagnetic order from first principle calculations \cite{boukhvalov2010stable}. In chapter \ref{chap:magnetization}, the origin of magnetic order in half-passivated graphene is discussed. Furthermore, the dependence of the magnetic order on the type of passivation will be explained.

In chapter \ref{chap:graphone}, first the two main issues for the fabrication of half-hydrogenated graphene and the proposed solution to resolve these issues will be addressed. Then in chapter \ref{chap:magnetization}, the origin of magnetization and magnetic properties of half-passivation graphene will be discussed. 

\chapter[HOW TO STABILIZE HALF-PASSIVATED GRAPHENE (GRAPHONE) IN EXPERIMENT]{\uppercase {HOW TO STABILIZE HALF-PASSIVATED GRAPHENE (GRAPHONE) IN EXPERIMENT}\footnote[1]{Part of the data reported in this chapter is reprinted with permission from my work in Ref. \cite{PhysRevB.90.035433}}} \label{chap:graphone}
In chapter \ref{chap:graphene}, the magnetization induced by graphene functionalization has been reviewed. It has been shown passivation of graphene with add-atoms like H and F breaks the sublattice symmetry and creates an imbalance in spin polarization belong to the different sublattices. Later in chapter \ref{chap:magnetization} it will be explained that the maximum magnetization is induced by passivation of the carbon atoms belonging to the single sublattice (depends on adatoms) where the resulting material is called graphone. Fundamentally, there are two main issues which hamper the fabrication of graphone in the experiment.
i) sublattice symmetry which results in the absence of preference sites for adatoms adsorption , and ii) lack of a migration barrier to trap adatoms and prevent them from migrating between the sublattices \cite{boukhvalov2010stable}.

Recently, there has been some progress in fabrication of partially hydrogenated-graphene \cite{peng2014new, giesbers2013interface} under certain experimental conditions (i.e low temperature) \cite{peng2014new}. Zhou et al proposed a functionalized heterostructures of fully hydrogenated graphene (graphane) on the top of hexagonal-boron nitride (h-BN) \cite{zhou2012fabricate}. The idea is creating an active nitrogen agent by exposing the system to fluorine. The instability of nitrogen-fluorine bond will increase the electronegativity of nitrogen thus creating an active nitrogen site. By applying pressure on the fluorinated h-BN layer, the system undergoes a structural transition from graphane to semi-hydrogenated graphene by adsorption of all the hydrogen atoms from one sublattice in graphane to h-BN layer \cite{zhou2012fabricate}. 

Hexagonal-boron nitride (h-BN) is the insulating isomorph of graphite (honeycomb lattice with boron and nitrogen on two adjacent sublattices) with a large band gap of 5.97 eV. It was shown \cite{dean2010boron} that h-BN is a superior substrate for graphene for homogenous and high quality graphene fabrication \cite{dean2010boron, kim2013synthesis, yang2013epitaxial}. The small lattice mismatching ($1.7$\%) and atomically planer structure of h-BN (free of dangling bonds and charge traps) preserve properties of graphene such as charge carrier mobility.

\section{Stacking Graphene on h-BN}
\label{sec:stacking}
In this section, I discuss the different ways of stacking in the graphene/h-BN layers, in which both graphene and h-BN have the hexagonal lattices. The graphene has carbon atoms in both sublattices, while h-BN has boron in one sublattice and nitrogen in the other.

\begin{figure}[h]
\centering
\centerline{\includegraphics[width=0.8\columnwidth]{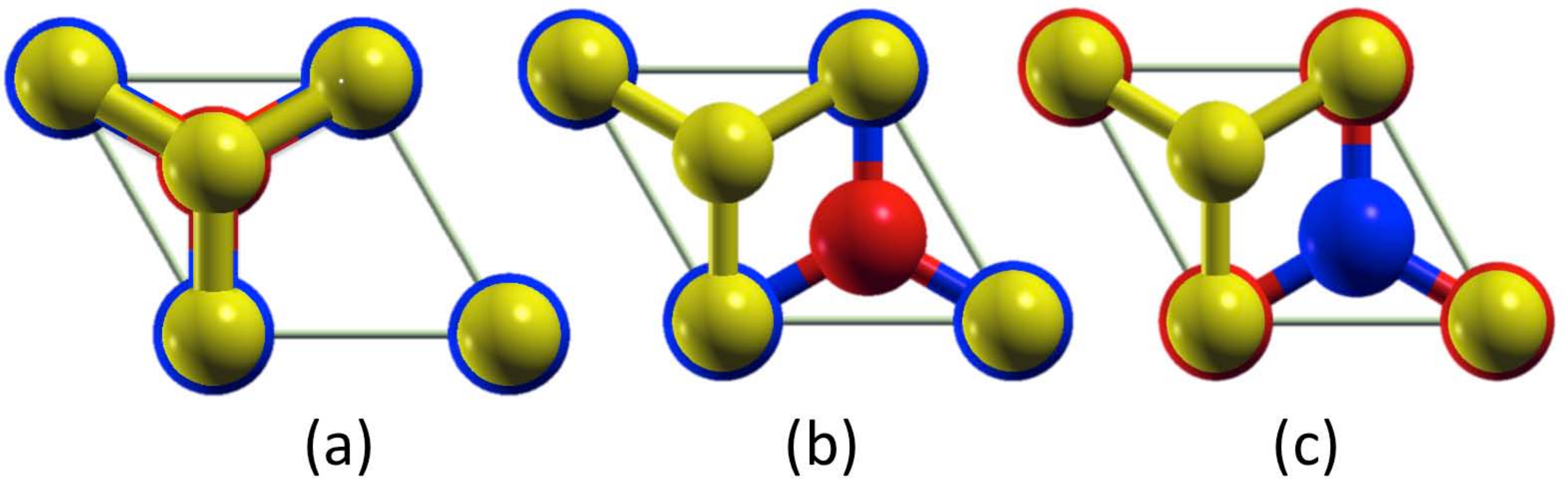}}
\centerline{\includegraphics[width=0.17\columnwidth]{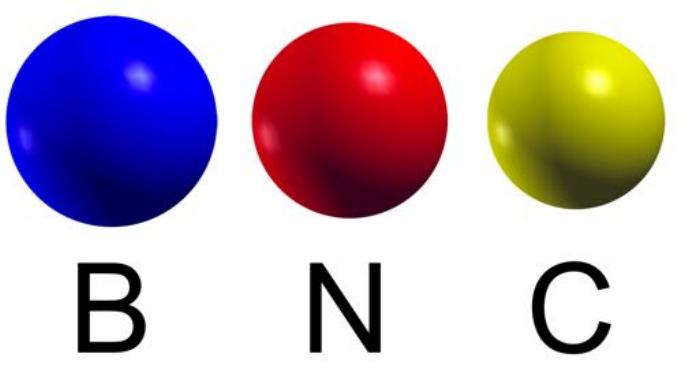}}
\caption{Different ways of stacking of graphene on top of h-BN: a) AA stacking; b) AB-I stacking; b) AB-II stacking. See full description in the text.} 
\label{fig:stack}
\end{figure}

Let us, at the moment, ignore the small (about 1.7 $\%$) mismatch between the graphene and h-BN lattices. The three most symmetric stacking variations are AA, AB-I, and AB-II: for AA stacking, (Fig. \ref{fig:stack}a). Each carbon atom of one sublattice of graphene is on top of a boron atom of h-BN, while each carbon atom of the other sublattice of graphene is on top of the nitrogen atom of h-BN; 
for AB-I staking, Fig. \ref{fig:stack}b, each carbon atom of one sublattice of graphene is on top of a boron atom of h-BN, while each carbon atom of the other sublattice of graphene is on top of the center of a hexagon of h-BN; for AB-I staking, Fig. \ref{fig:stack}c, each carbon atom of one sublattice of graphene is on top of nitrogen atom of h-BN, while each carbon atom of the other sublattice of graphene is on top of the center of a hexagon of h-BN. First principle calculations \cite{PhysRevB.76.073103,slawinska2010energy} showed that AB-II stacking configuration has a much larger energy and thus can be ignored. For the rest of the thesis I use AB to refer to AB-I stacking.

Although, the precise stacking control of graphene on h-BN is problematic, Wei Yang {\it et. al.} \cite{yang2013epitaxial} showed  that epitaxial growth of graphene on h-BN may allow pure selectional stacking in this heterostructure. 

Different stacking types affect the electrical properties of the substrate, such as the dipole moment differently. AB stacking provides larger asymmetry between boron and nitrogen environments of h-BN lattice than AA stacking. This difference is taken into account in the calculations.

\begin{figure}
\centering
\includegraphics[width=0.8\columnwidth]{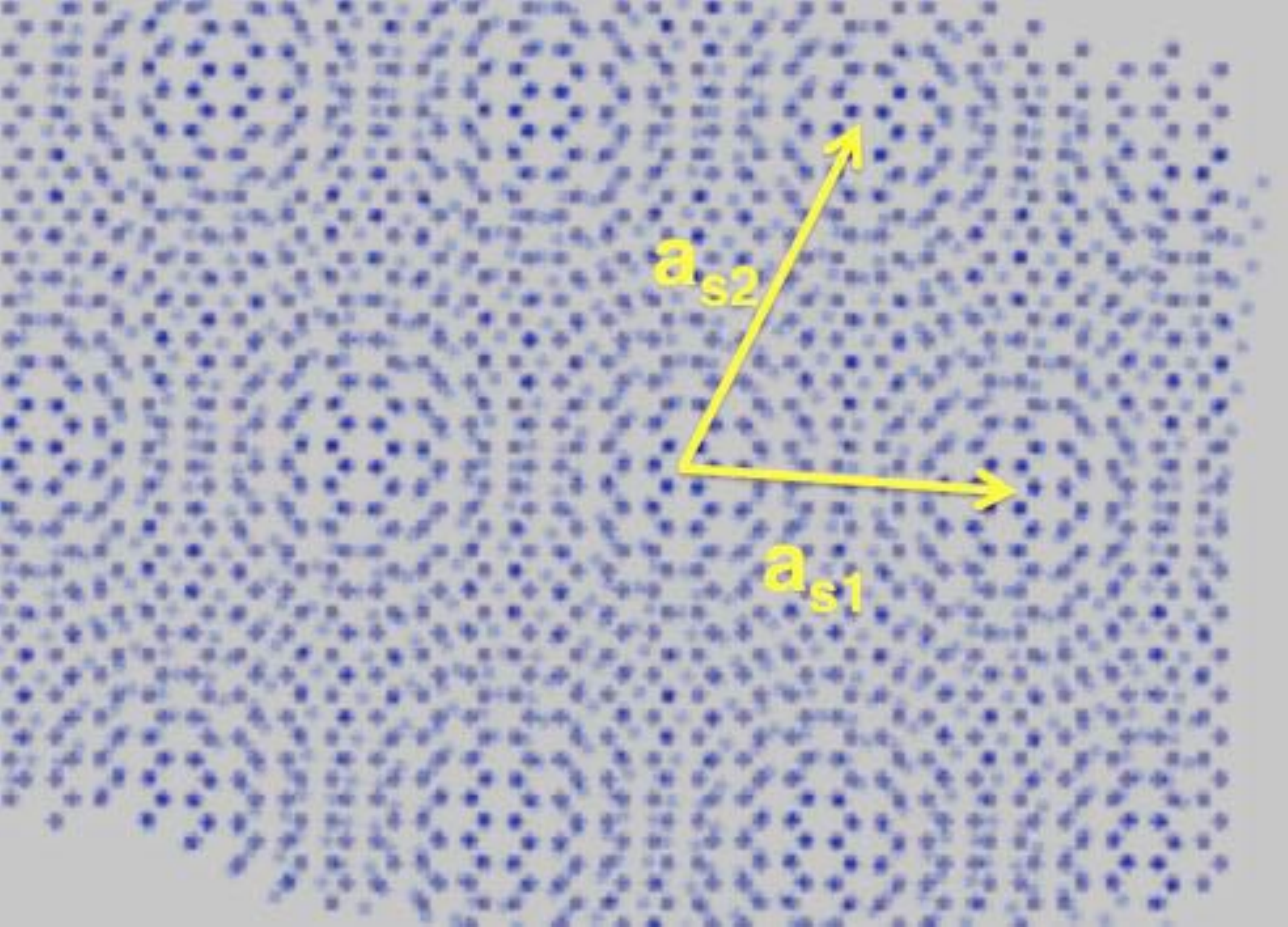}
\caption{Creation of moir\'e pattern in the presence of lattice mismatching between graphene and h-BN sublattice. The vectors \textbf{$a_{s_1}$}  and \textbf{$a_{s_2}$} are the supercell lattice vectors.} 
\label{fig:flower}
\end{figure}

Consider now a small lattice mismatching (1.7 $\%$) between graphene and h-BN lattices. This mismatch will create a large moir\'e pattern as seen in Fig. \ref{fig:flower}. Given small lattice mismatching and small twisting angle (in the absence of pure AB or AA-stacking), the supercell  can be divided into two main stacking domains: AA-stacking at the corners and AB-stacking at 1/3 and 2/3 of the long diagonal \cite{trambly2010localization}. 

These domains fill most of the area inside the supercells. For each domain, there is  a preference site for hydrogen adsorption on just one sublattice related to the stacking of the domain.

\section{Substrate Effect of h-BN to Stabilize Half-Hydrogenated Graphene}\label{sec:hydrogenation}
In this section, a practical method to solve the two obstacles (symmetry of two sublattices and mobility of hydrogen) using the functionalized graphene hybrid structure with h-BN is presented. The proposed experimental set up includes two steps: fabrication of graphene on h-BN substrate then exposing the system into the hydrogen plasma. The electrical dipole induced by the substrate in addition to small buckling of carbon bonds will trap hydrogen in one sublattice and will kinematically stabilize the system.

I show that for h-BN the difference in electronegativity of nitrogen and boron in h-BN creates a dipole moment for each nitrogen site. This dipole moment breaks the equivalency of two carbon atoms in two different graphene sublattices. The similar screening effect has been reported in multilayer graphene but with different strength  \cite{PhysRevB.91.155419}. Moreover, the screening effect of h-BN will generate a buckling in the graphene layer. This buckling will change the vertical position of the one sublattice with respect to the other sublattice and will enhance the coverage rate of the hydrogen in one sublattice. The dipole moment is also responsible for the increased migration barrier in the adsorbed hydrogen atoms, effectively pinning the hydrogen atoms to one sublattice.

\subsection{Hydrogen Adsorption}
\label{sec:stability}
To indicate the substrate effect on selective hydrogen adoption in single sublattice, I first show that in the graphene h-BN heterostucture for both stacking AA and AB, the hydrogen is adsorbed predominantly on one sublattice of the graphene lattice. The half hydrogenated graphene then normally becomes perfect graphone. I show that this graphone is stable with respect to hydrogen migration and desorption. I compare the energy differences, migration barriers, migration energies and binding energies of AA and AB stacking with that of pristine graphone.        

\begin{figure}[h]
\centering
\centerline{\includegraphics[width=0.8\columnwidth]{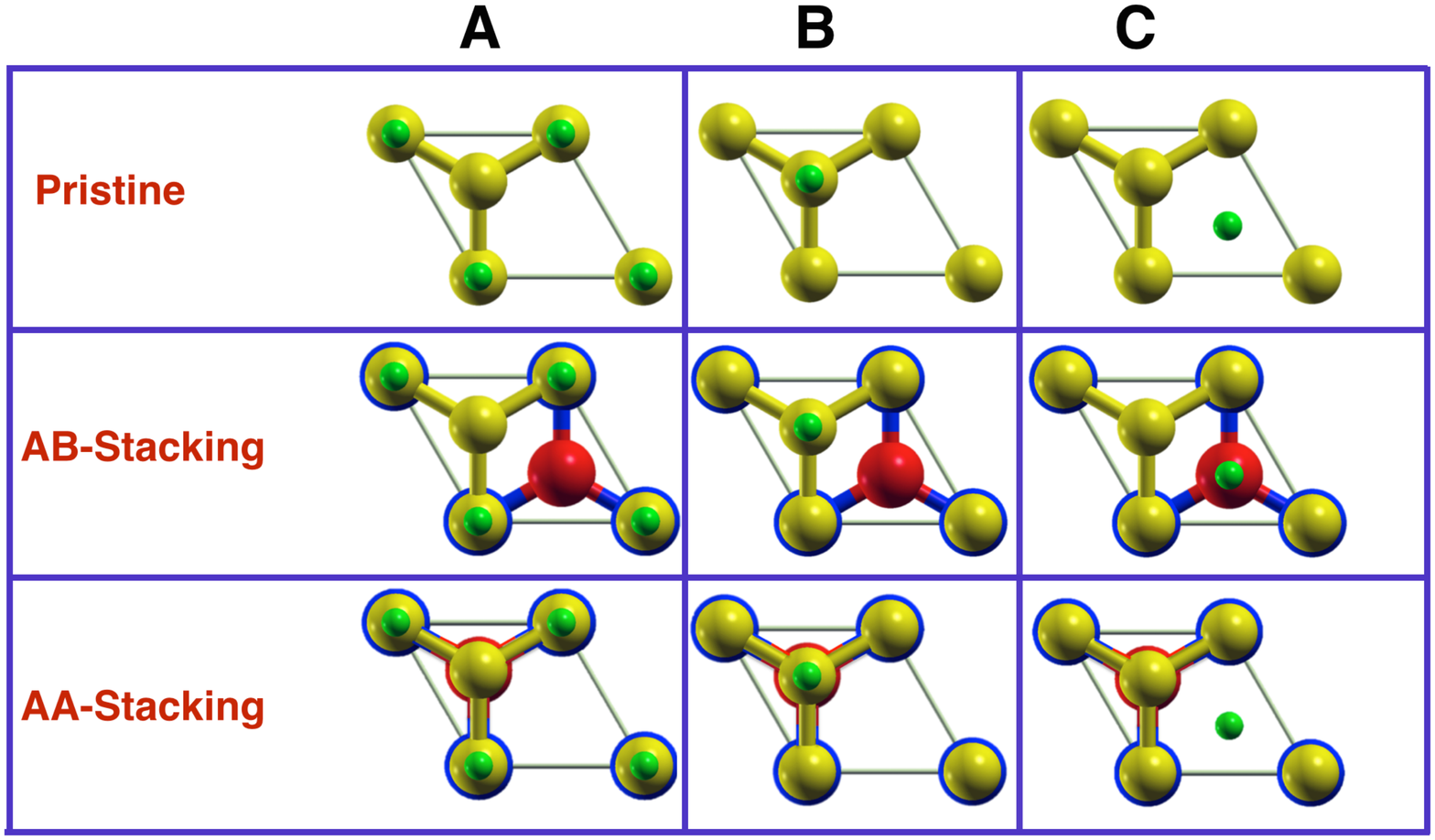}}
\centerline{\includegraphics[width=0.15\columnwidth]{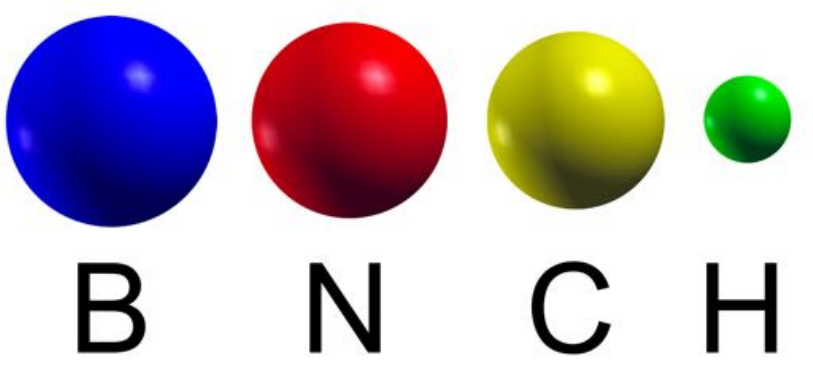}}
\caption{The first row shows different hydrogenations of graphene, the second row shows  different hydrogenations of AB stacking, and the third shows that for AA stacking. Hydrogenation differs by the placement of the hydrogen atom within the unit cell. The first column is graphone(A), the second is graphone(B) and the third is graphone(C). I refer to them in the text as AB-C -- this is AB stacked heterostructure which forms graphone(C).}
\label{fig:2}
\end{figure}

There are three high symmetry adsorption sites per unit cell for pristine graphone and for any stacking types of the heterostucture. The hydrogen atom can be adsorbed on the sublattice A, sublattice B, or in the center of the hexagon C of graphene, see Fig \ref{fig:2}. 

I use LDA, GGA, and VDW methods to find the energy difference, $\Delta E_{\mbox{\textsc{ba}}}=E_{\mbox{\textsc{a}}}-E_{\mbox{\textsc{b}}}$ and $\Delta E_{\mbox{\textsc{bc}}}=E_{\mbox{\textsc{c}}}-E_{\mbox{\textsc{b}}}$, between different graphones, A, B, and C, for both stacking types and for the pristine case. The results were shown  in the Table \ref{table:1}.   

\begin{table}[ht]
\centering
\begin{tabular}{| m{1.40cm} | m{3.2cm} | m{3.2cm} |}
\hline
Config.&\multicolumn{1}{c|}{$\Delta E_{\mbox{\textsc{ba}}}$}&\multicolumn{1}{c|}{$\Delta E_{\mbox{\textsc{bc}}}$} \\
\hline
&LDA GGA VDW&LDA GGA VDW\\
Pristine& \multicolumn{1}{c|}{0}& \multicolumn{1}{c|}{0.02}\\
AB&0.02 \space\space 0.01 \space\space 0.03 &0.03 \space\space 0.02 \space\space 0.05\\
AA&0.01 \space\space0.02 \space\space 0.02 & 0.03 \space\space 0.07 \space\space 0.08\\
\hline
\end{tabular} 
\caption{Energy differences (in eV) per unit cell, $\Delta E_{\mbox{\textsc{ba}}}=E_{\mbox{\textsc{a}}}-E_{\mbox{\textsc{b}}}$ and $\Delta E_{\mbox{\textsc{bc}}}=E_{\mbox{\textsc{c}}}-E_{\mbox{\textsc{b}}}$ for different types of graphone, A , B, C, pristine graphone and for different kinds of stacking AB and AA. For three different types of configurations (AA, AB, and pristine graphone) using LDA, GGA, and VDW approximations.} 
\label{table:1}
\end{table} 

For pristine graphone, where there are no inter-planar VDW interactions, I have reproduced the results of the previous work \cite{boukhvalov2010stable} which used GGA calculations. It is also no surprise that the graphone(A) and graphone(B) have the same energy, as in the pristine case there is no difference between them. The energy of a pristine graphone(C) is larger than the energy of pristine graphone(A) or (B).

For the graphene h-BN heterostructure the results were different. In both AA and AB stacking there is no symmetry between different sublattices. From Table \ref{table:1}, it is seen that the graphone(B) where hydrogen adsorbs to the sublattice B (the sublattice where is not on top of the boron atoms for both AA and AB stacking types), has the smallest energy for both ways of stacking. All three methods of LDA, GGA, and VDW also show that for both kinds of stacking, graphone(C) has a larger energy than graphone(A).

The conclusion is that the site shown in the column B of Fig. \ref{fig:2} is the preference site for hydrogen adsorption. If the hydrogen atoms adsorb to all sites B on the hexagonal lattice, the result is pure graphone(B). 

Later the stability of this graphone will be checked. First, the graphone(B) for both ways of stacking against removal of one hydrogen atom or two neighboring hydrogen atoms is checked. The results of LDA, GGA, and VDW calculations for both kinds of stacking as well as for pristine graphone were summarized in Table \ref{table:3}.   

\begin{table}[ht]
\centering
\begin{tabular}{ | m{1.4cm} | m{3.2cm} | m{3.3cm} | }
\hline
Config.&H-Binding energy&Removal of two H\\
\hline
&LDA GGA VDW&LDA GGA VDW\\
Pristine& \multicolumn{1}{c|}{1.15} & \multicolumn{1}{c|}{-5.31}\\
AB&1.10 \space\space 1.01 \space\space 1.12&0.32 \space\space 0.36 \space\space 0.38\\
AA&1.09 \space\space 1.01 \space\space1.10& 0.31 \space\space 0.33 \space\space 0.34\\
\hline
\end{tabular} 
\caption{ Binding energy for one and two neighboring hydrogen atoms  for graphone(B) for pristine, AA, and AB stacking variations. The energies are given in eV per unit-cell.} 
\label{table:3}
\end{table} 

The conclusion is that graphone(B) is stable against desorption of hydrogen for both AA and AB stacking types of the h-BN/graphene heterostructure. It is noted that the pristine graphone is not stable against desorption of two neighboring hydrogen atoms.  

Finally, I check the stability of graphone(B) against migration of the hydrogen atom to a nearest site on another sublattice. In order to do that, I calculate both the migration energy and migration barrier. I denote these graphone types as graphone(A), graphone(B) and graphone(C).

For finding the migration energy, one hydrogen atom has been moved per supercell ($5\times5$ unit-cell) to the nearby sublattice and then the energy difference per unit-cell for these two configurations  has been found. The migration barrier has been derived by the minimum-energy path calculation under the NEB method.\cite{PhysRevB.80.085428} NEB results in four different supercells, $2\times2$
, $3\times3$, $4\times4$ and $5\times5$ indicate a fairly small (less than 5 $\%$) correlation effect of two distorted adjacent supercells. For pristine graphone, only the GGA calculation has been performed and results are in agreement with the previous work. \cite{boukhvalov2010stable}

\begin{table}[ht]
\centering
\begin{tabular}{ |m{2cm} | m{2.5cm} | m{3.3cm} | }
\hline
Config.&Migration barrier&Migration energy\\
\hline
        &\multicolumn{1}{c|}{  LDA (NEB)}&LDA GGA VDW\\
Pristine&\multicolumn{1}{c|}{0.06}&\multicolumn{1}{c|}{-1.44} \\
AB      &\multicolumn{1}{c|}{0.18}&-0.87 \space-1.04 \space -0.64\\
AA      &\multicolumn{1}{c|}{0.12}&-0.93 \space -1.12 \space-0.82 \\
\hline
\end{tabular} 
\caption{Migration barrier and migration energy for three different types of configurations.} 
\label{table:4}
\end{table} 

It has been previously shown \cite{boukhvalov2010stable} that the small migration barrier makes the pristine graphone unstable against hydrogen migration. Table \ref{table:4} shows that this barrier is just $0.06$ eV. The table also shows a substantial increase in the migration barrier for both AA and AB stacking. This increase is due to the screening effect of h-BN.

Such barriers decreases the mobility of hydrogen atoms on top of the graphene layer in the presence of h-BN. If I start with half hydrogenated graphene, from the Arrhenius equation, $e^{(-\frac{\Delta E}{k_{B}T})}$, at room temperature, the transition probability will be less than 0.001 and from $\tau \sim \frac{\hbar}{k_{B}T}e^{(\frac{\Delta E}{k_{B}T})}$ the naive estimate for transition time will be $~10^{-10}$ sec. The transition time increases rapidly by decreasing temperature whereas at $T=60 K$, the transition time reaches to the order of 1 sec.

\begin{figure}[!h]
\centering
\includegraphics[width=0.4\columnwidth]{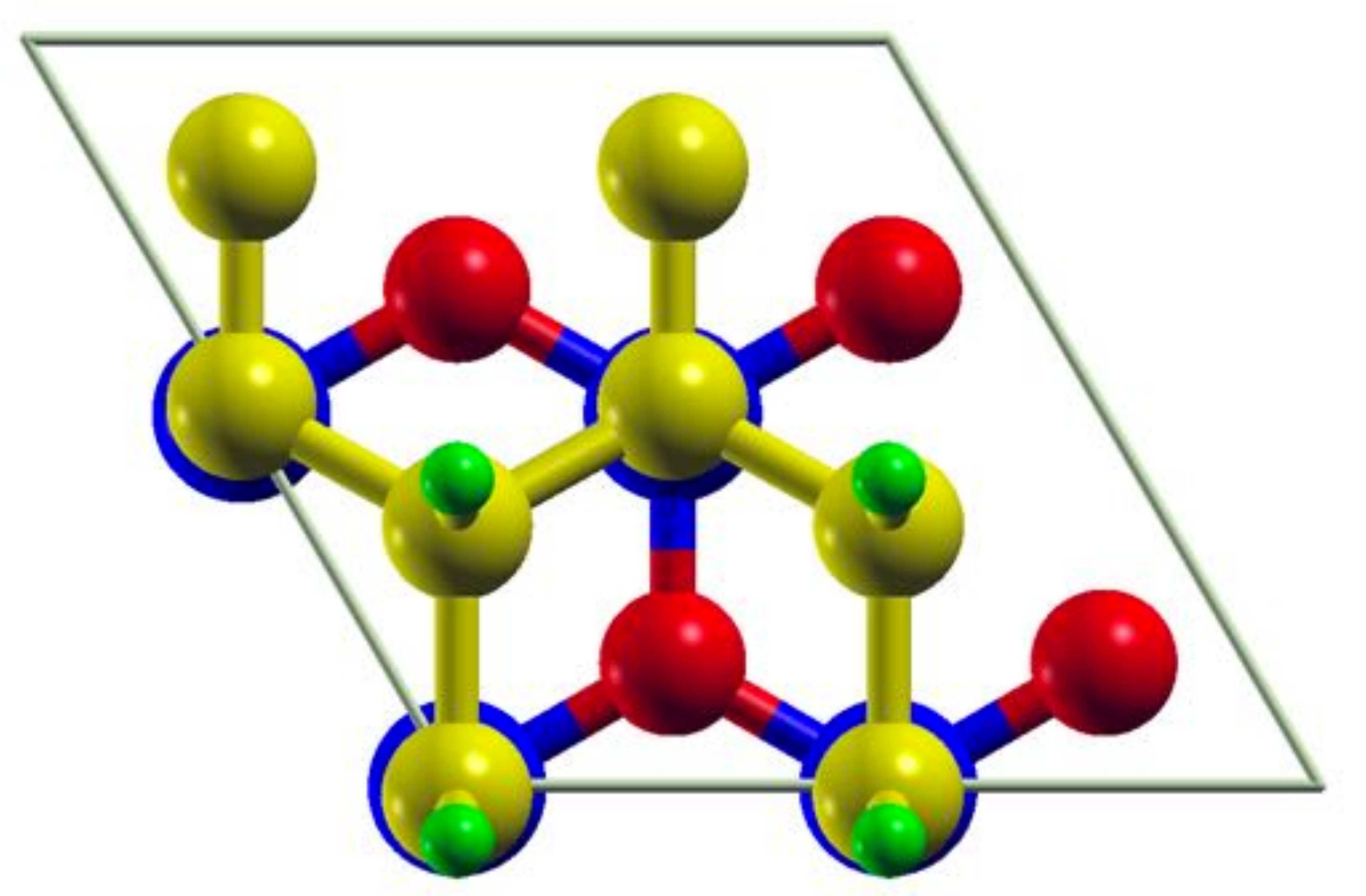}
\includegraphics[width=0.2\columnwidth]{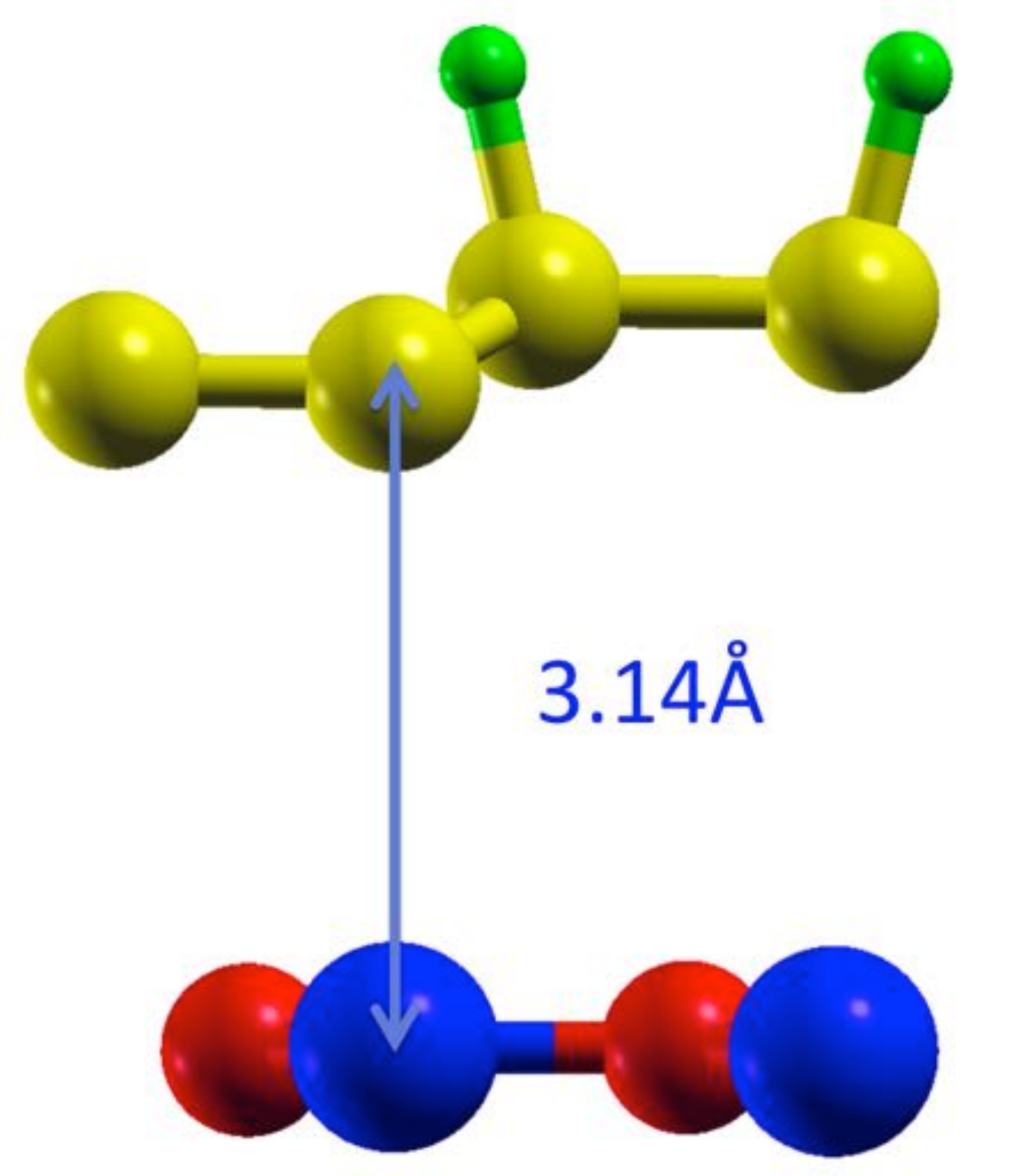}
\caption{Optimized structure for boat graphone/h-BN with 2$\times$2 supercell.}
\label{fig:boat}
\end{figure}

The migration energy for both types of stacking is still negative, although substantially less than for  pristine graphone. Although this still makes the boat graphone (see Fig. \ref{fig:boat}) the most stable configuration thermodynamically, if one exposes the graphene/hBN to the hydrogen plasma, graphone(B) will be the most stable case kinematically. This can be understood from the fact that carbon atoms belonging to sublattice B come first in contact with hydrogen atoms and they then encounter a large migration barrier due to the presence of the substrate. Moreover, it has been shown recently \cite{kharche2011quasiparticle, pumera2013graphane} that half-hydrogenated graphene only in one sublattice can be fabricated within the selective desorption of hydrogen atoms from one side of the chair graphone \cite{PhysRevB.75.153401}. Such produced graphone then will be stabilized by the high migration barrier for the h-BN/graphene heterostructure.

\section{Substrate Effect of h-BN to Stabilize Half-Fluorinated Graphene}
In previous section, the technical difficulties to stabilize hydrogen atoms on graphene in a controlled pattern have been discussed. Moreover, since the binding energy of hydrogen molecule is relatively large, breaking the hydrogen bond followed by adsorption of hydrogen atoms on graphene is endothermic and unfavorable. Therefore, hydrogenation of graphene in a single sublattice requires using hydrogen plasma. More importantly, in previous chapter I prove the necessity of substrate like h-BN in order to control passivation of hydrogen atoms in a single sublattice. 

However, the relatively large formation energy of half-fluorinated graphene compared to half-hydrogenated graphene \cite{boukhvalov2010stable} enables formation of fluorinated-graphene even from fluorine molecules ($F_2$). This can be seen also from small dissociation energy of $F_2$ (159 $KJ mol^{-1}$) molecules compared to $H_2$ (458 $KJ mol^{-1}$). 

Comparing migration energy of fluorine (-1.44 eV) with hydrogen adatoms (0.17 eV) in two different graphone systems functionalized with hydrogen and fluorine, respectively shows that half-fluorinated graphene in a single sublattice is more stable.  

However, recent studies show that even achieving half-fluorinated graphene on one side is a big challenge \cite{boukhvalov2016absence}. Significant distortion of graphene flake during fluorination creates a meta-stable pattern and prevents further fluorination process. In this thesis, application of  h-BN to prevent passivation on both sides and to reduce distortion of the graphene flacks caused by passivation.

As discussed in the previous section (\ref{sec:hydrogenation}), the screening effect of h-BN breaks the sublattice symmetry and traps adatoms from moving to the other sublattice thus stabilizes passivation in a single sublattice. \cite{cernov2014screening} 
 
 To show the effect of h-BN on stabilizing half-fluorinated graphene, similar calculations proposed in section \ref{sec:hydrogenation} were followed.
 
Following by the same method discussed in section \ref{sec:hydrogenation} I have performed the density functional calculations using LDA, GGA, and VDW for the exchange-correlation functional approximation to find the energy difference, $\Delta E_{\mbox{\textsc{ba}}}=E_{\mbox{\textsc{a}}}-E_{\mbox{\textsc{b}}}$ and $\Delta E_{\mbox{\textsc{bc}}}=E_{\mbox{\textsc{c}}}-E_{\mbox{\textsc{b}}}$, between different half-fluorinated graphene, A, B, and C, for both stacking types and for the pristine case. The results are shown in the Table \ref{table:f1}.   

\begin{table}[ht]
\centering
\begin{tabular}{| m{1.8cm} | m{3.6cm} | m{3.6cm} |}
\hline
Config.&\multicolumn{1}{c|}{$\Delta E_{\mbox{\textsc{ba}}}$}&\multicolumn{1}{c|}{$\Delta E_{\mbox{\textsc{bc}}}$} \\
\hline
& \ LDA GGA VDW& \ LDA \ GGA \ VDW\\
Pristine& \multicolumn{1}{c|}{0}& \multicolumn{1}{c|}{0.546}\\
AB&0.011 \space\space 0.009 \space\space 0.014 &0.061 \space\space 0.055 \space\space 0.064\\
AA&0.010 \space\space 0.008 \space\space 0.012 & 0.082 \space\space 0.076 \space\space 0.088\\
\hline
\end{tabular} 
\caption{Energy differences (in eV) per unit cell, $\Delta E_{\mbox{\textsc{ba}}}=E_{\mbox{\textsc{a}}}-E_{\mbox{\textsc{b}}}$ and $\Delta E_{\mbox{\textsc{bc}}}=E_{\mbox{\textsc{c}}}-E_{\mbox{\textsc{b}}}$ for different types of graphone, A , B, C, pristine graphone ($C_2F$) and for different kinds of stacking AB and AA. For three different types of configurations (AA, AB, and pristine graphone) using LDA, GGA, and VDW approximations.} 
\label{table:f1}
\end{table}

 For pristine graphone, the results of the previous work \cite{boukhvalov2010stable} which used GGA calculation are reproduced. For the graphene h-BN heterostructure the results are different. In both AA and AB stacking there is no symmetry between different sublattices. 
 
 From Table \ref{table:f1}, I see that the graphone(B) where fluorine adsorbs to the sublattice B (the same as half-hydrogenated graphene presented in Fig. \ref{fig:2}), has the smallest energy for both ways of stacking. I conclude that the site shown in the column B of Fig. \ref{fig:2} is the preference site for fluorine adsorption which is also the case for half-hydrogenated graphene.

\begin{figure}
\centering
\includegraphics[width=0.8\columnwidth]{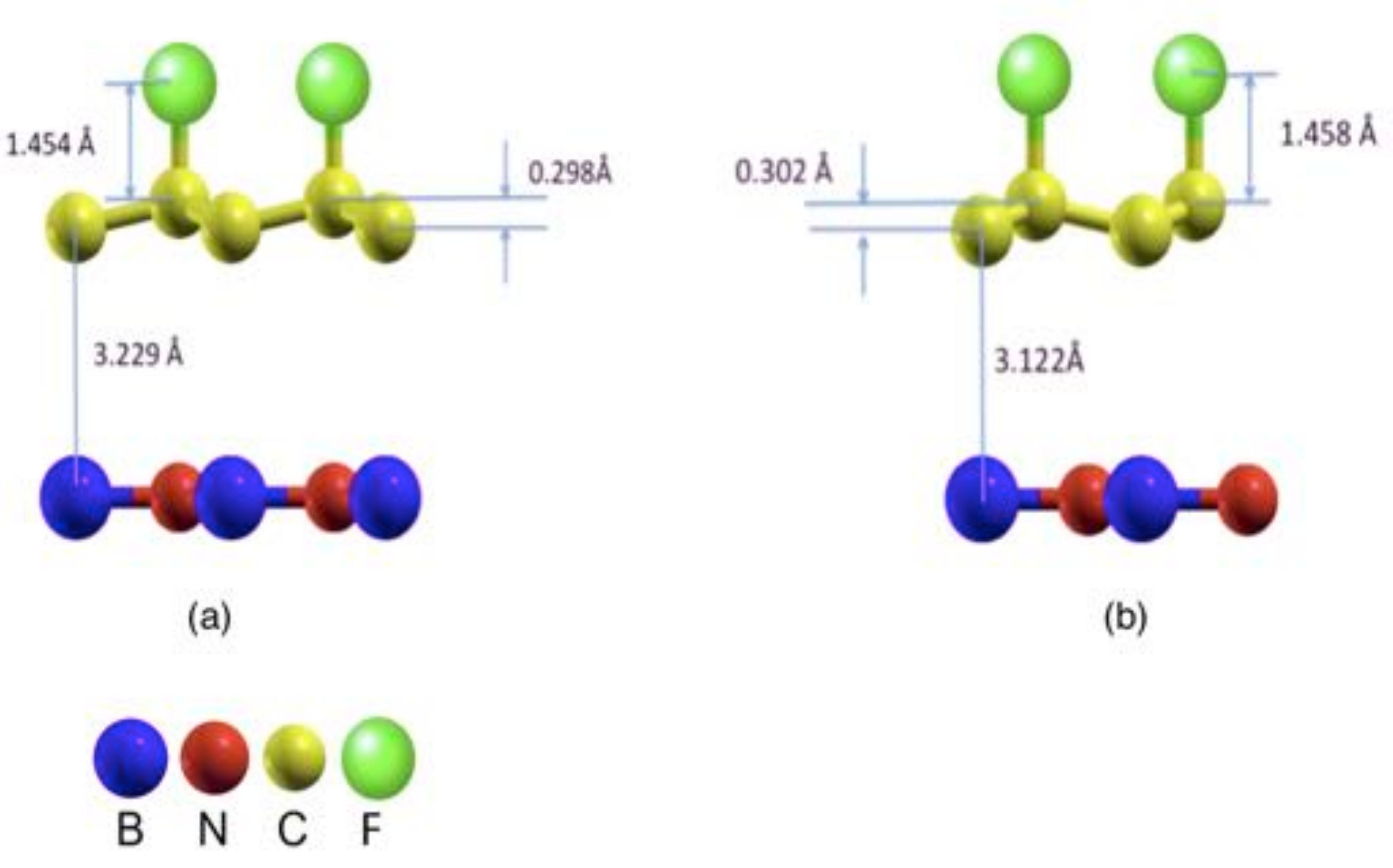}
\caption{Optimized geometry of half-fluorinated graphene: on the top of h-BN: (a) For AA-stacking, (b) AB stacking.} 
\label{fig:c2fgeo}
\end{figure}

Fig. \ref{fig:c2fgeo} shows the optimized geometry of half-fluorinated graphene on the top of h-BN for both AA-stacking (Fig. a) and AB stacking (Fig. b). I have carried out LDA, GGA and VDW calculations to obtain the vertical distance between carbon and boron atoms ($d_{CB}$) and the vertical distance between two nearest-neighbor carbon atoms ($d_{CC}^{\perp}$). The results has been summarized in Table \ref{table:f2}.
 
\begin{table}[!h]
\centering
\begin{tabular}{|c|cc|cc|cc|}
\hline 
Structure & LDA& & GGA & & VDW & \\
&$d_{CB}$ & $d_{CC}^{\perp}$ &$ d_{CB} $&$ d_{CC}^{\perp}$ &$ d_{CB}$ &$ d_{CC}^{\perp}$ \\ 
\hline
$C_2F$/h-BN (AA-stacking)& 3.22 & 0.30 & 3.23 & 0.30 & 3.23 & 0.30 \\ 
$C_2F$/h-BN (AB-stacking) & 3.20 & 0.30 & 3.13 & 0.31 & 3.13 & 0.31 \\ 
\hline
\end{tabular} 
\caption{Optimized structure (in \AA): $d_{CB}$ is the distance between carbon on top of the boron atom and $d^{\perp}_{cc}$ is the vertical distance between two carbon atoms which belong to the same layer. It shows the magnitude of the bond bending.} 
\label{table:f2}
\end{table}

To check the stability of half-fluorinated graphene on h-BN, I consider graphone(B) for both ways of stacking (AA and AB) against removal of one fluorine atom or two neighboring fluorine atoms. The results of LDA, GGA, and VDW calculations for both kinds of stacking as well as for pristine graphone are shown in Table \ref{table:f3}.   

\begin{table}[ht]
\centering
\begin{tabular}{ | m{1.4cm} | m{3.2cm} | m{3.3cm} | }
\hline
Config.&F-Binding energy&Removal of two F\\
\hline
&LDA GGA VDW&LDA GGA VDW\\
Pristine& \multicolumn{1}{c|}{1.15} & \multicolumn{1}{c|}{0.50}\\
AB&1.14\space\space 1.14 \space\space 1.14&2.44 \space\space 2.41 \space\space 2.68\\
AA&1.14 \space\space 1.13 \space\space1.14& 2.23 \space\space 2.18 \space\space 2.35\\
\hline
\end{tabular} 
\caption{Binding energy for one and two neighboring fluorine atoms  for graphone(B) for pristine, AA, and AB stacking variations. The energies are given in eV per unit-cell.} 
\label{table:f3}
\end{table} 

The conclusion is that $C_2F$(B) is even more stable against desorption of fluorine for both AA and AB stacking types of the h-BN/graphene heterostructure compared to pristine $C_2F$. In addition, due to the screening effect of h-BN, removing two fluorine atoms costs more energy in the presence of h-BN. This shows h-BN reduces distortion effect caused by fluorine adsorption which was the main issue reported in the literature. \cite{boukhvalov2016absence}

\chapter[MAGNETIC PROPERTIES OF HALF-PASSIVATED GRAPHENE]{\uppercase{MAGNETIC PROPERTIES OF HALF-PASSIVATED GRAPHENE} \footnote{Part of the data reported in this chapter is reprinted with permission from my work in Ref. \cite{PhysRevB.90.035433}} }\label{chap:magnetization}
One of the main reasons why pristine graphone(A) or graphone(B) are interesting and promising is the magnetic properties of these materials.  It has been shown by spin-polarized first principle calculation in LSDA scheme  that graphone (A or B) has about 1 $\mu_{B}$ per unit-cell magnetization \cite{zhou2009ferromagnetism}. Unfortunately, pristine graphone is not stable enough to be of any practical use.  

In the previous chapter \ref{chap:graphone}, it has been seen that the h-BN substrate considerably increases the graphone(B) stability, making it feasible for further studies. It is, however, important to show that such stabilized graphone still has the magnetic properties expected from the pristine graphone. 

In this chapter, the origin of magnetization in half-hydrogenated and half fluorinated graphene has been described. Then the results of the LSDA with GGA calculations for graphone(B) on h-BN substrate are presented.

\section{Magnetization Root in Graphone}
\begin{figure}[!h]
\centering
\includegraphics[width=0.8\textwidth]{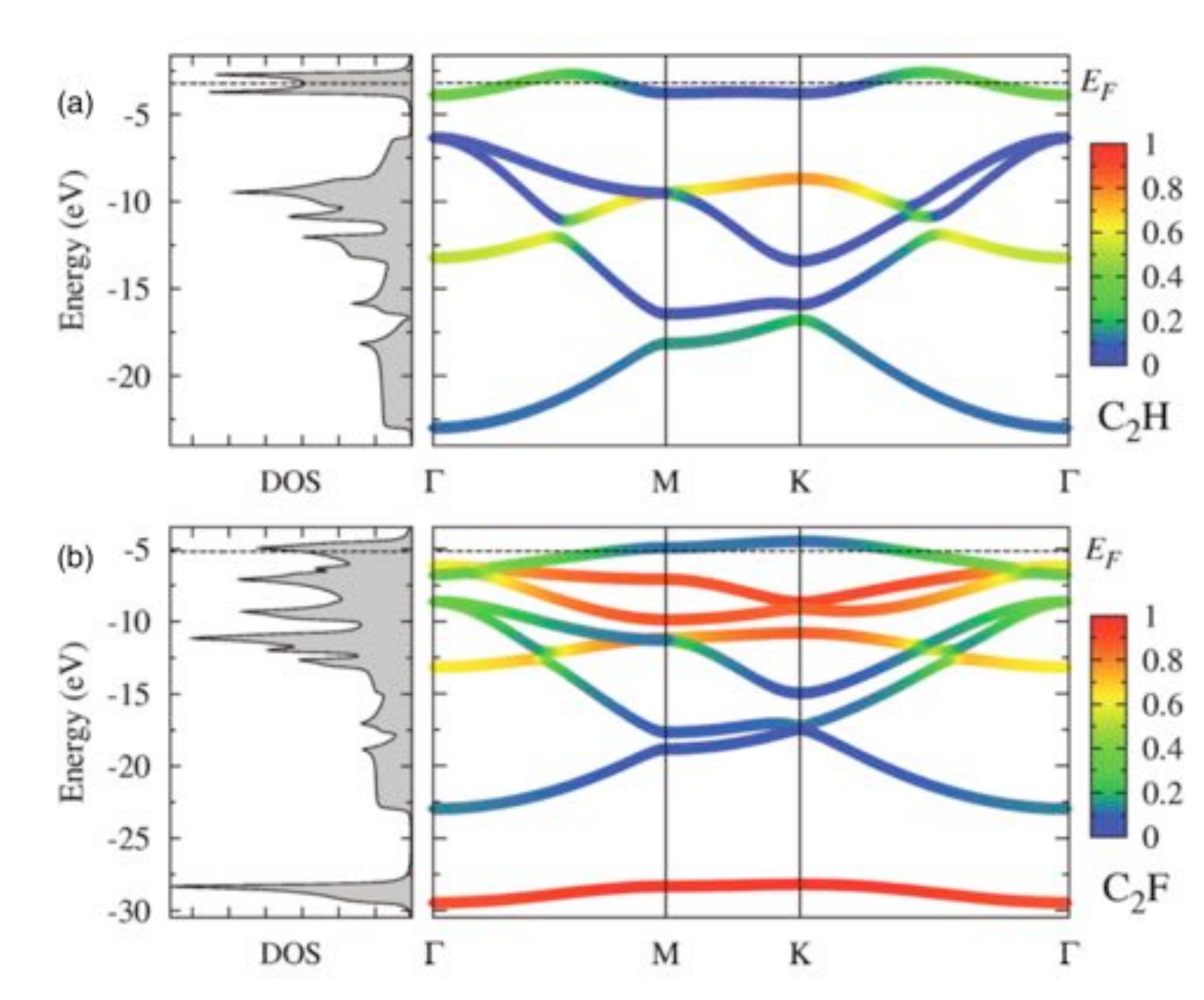}
\caption{DOS (left) and band-structure (right) in a) half-hydrogenated graphene ($C_2H$) and b) half-fluorinated graphene ($C_2F$). The band-structure (right) of both $C_2H$ and $C_2F$ have been projected into the impurity orbitals ($s$-orbital for $C_2H$  and $sp$-orbital for $C_2F$). Figure is reprinted with permission from Ref. \cite{PhysRevB.88.081405}. } 
\label{fig:bschcf}
\end{figure}

Magnetization of $sp$-materials both fundamentally and practically important particularly because of their relatively high currie-temperature \cite{edwards2006high}. In section \ref{sec:defect}, the origin of magnetization in graphene with a single defect has been discussed. In contrast to a single defect, magnetization in half-passivated graphene is missing a consistent fundamental description. Fig. \ref{fig:bschcf} shows the density of states (DOS) and band structure of half-hydrogenated graphene ($C_2H$) and half-fluorinated graphene ($C_2F$) \cite{PhysRevB.88.081405}. To calculate the contribution of impurity orbitals ($s$ for hydrogenated and $sp$ for fluorinated) Rudenko et.al. \cite{PhysRevB.88.081405} have utilized maximally-localized Wannier function \cite{RevModPhys.84.1419} (MLWF) to project the band structure of both $C_2H$ and $C_2F$ into impurity orbitals and the results has been shown in Fig. \ref{fig:bschcf}.

 Fig. \ref{fig:bschcf} shows for both $C_2H$ and $C_2F$, the density of states at the Fermi-energy is non-zero. The none zero density of states at fermi-energy mainly originated from unpaired electron localized at $p_z$ orbital of carbon together with some contribution from the impurity orbitals ($s$ in the case of $C_2H$ and $sp$ for $C_2F$). \cite{PhysRevB.88.081405}
 
 In Fig. \ref{fig:bschcf}  the contribution of the impurity orbital from the projected band-structure can be seen. The band-structure shows that in the case of $C_2H$ the impurity orbitals are mainly unfilled compared to $C_2F$. In $C_2F$, the $sp$-orbital (red color) are mainly filled thus the impurity orbital in fluorinated graphene plays the role of intermediate filled-orbital in the super-exchange model. 
 
\begin{figure}[H]
\centering
\centerline{\includegraphics[width=1.1\textwidth]{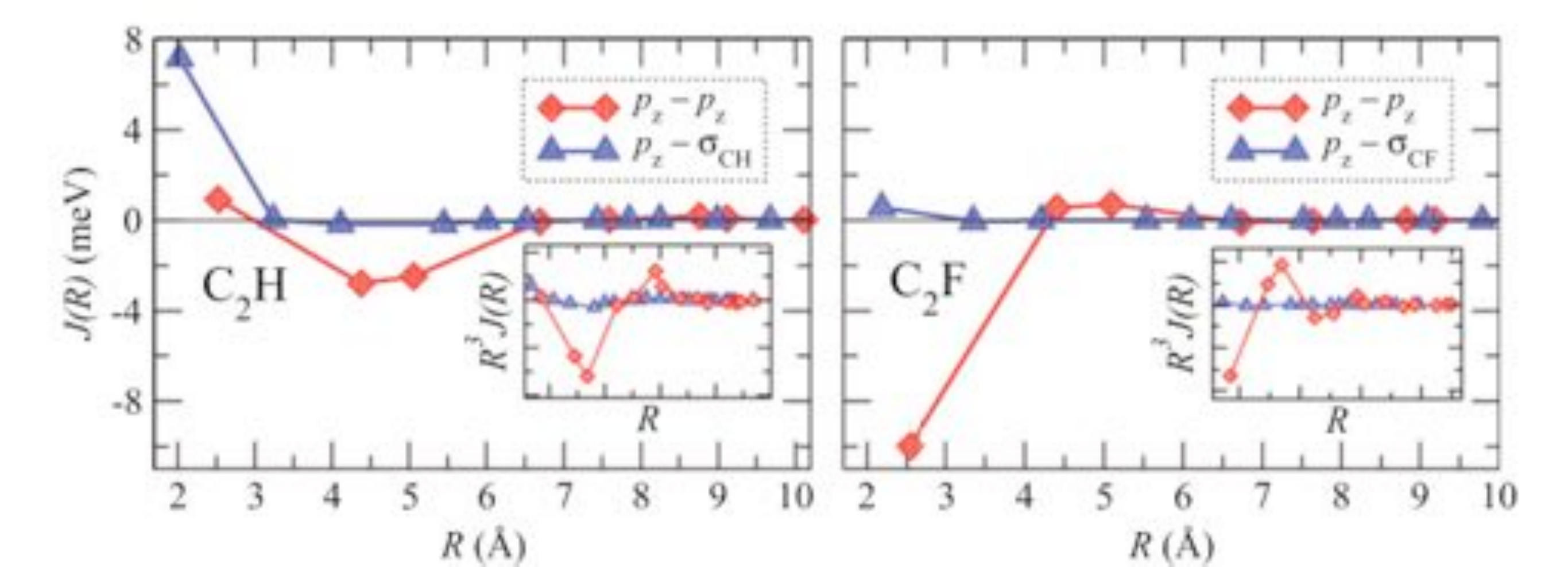}}
\caption{The exchange parameters as a function of distance for $C_2H$ (left panel) and $C_2F$ (right panel). In the left (right) panel the red and blue curves show the exchange parameter between $p_z-p_z$ orbitals and $p_z-\sigma_{CH}$ ($p_z-\sigma_{CF}$)  in $C_2H$ ($C_2F$), respectively. Figure is reprinted with permission from Ref. \cite{PhysRevB.88.081405}.}
\label{fig:J}
\end{figure}

Fig. \ref{fig:J} indicates  the exchange parameters between different pair of orbitals (with non-negligible contribution in exchange interaction) $p_z-p_z$, $p_z-\sigma_{CH(CF)}$ for both $C_2H$ and $C_2F$ as a function of distance. The dominant exchange parameter is positive for $C_2H$ and negative for $C_2F$ and this shows why these materials have different magnetization order. The dominant exchange parameter in $C_2H$ belongs to $p_z-\sigma_{CH}$ orbital pair. However, the exchange between $p_z-p_z$ orbital pair is the leading term in $C_2F$ and is negative. In $C_2H$ the sign of next leading term is changing (AFM coupling) and it belongs to $p_z-p_z$ orbital pairs. On the other hand, in $C_2F$, exchange parameter is negligible for the larger distances. \cite{PhysRevB.88.081405}

The sign of exchange parameter in $p_z-p_z$ orbital pair is changing in both $C_2H$ and $C_2F$ but it does not follow RKKY interaction \cite{fazekas1999lecture} since the interaction term decays faster than $R^{-3}$. Unsaturated $p_z$-orbitals in  $C_2H$ are mainly responsible for magnetic moments. However, ferromagnetic exchange coupling are mainly arising from FM-exchange interaction between unpaired $p_z$ orbital and impurity orbitals $\sigma_{CH}$. The difference between $C_2H$ and $C_2F$ arises from the difference in overlap between $p_z$  and $\sigma_{CH(CF)}$ orbitals \cite{PhysRevB.88.081405}.

\subsection{Super-Exchange V.S. Double-Exchange}
From the hopping parameters between different orbitals \cite{PhysRevB.88.081405}, it can be seen the difference between $C_2H$ and $C_2F$ magnetic order also originated from indirect-exchange. The indirect exchange arises from the presence of intermediate impurity orbitals ($\sigma_{CH(CF)}$). In $C_2F$ the impurity orbital is filled (see the band structure of $C_2F$ in Fig \ref{fig:bschcf}). Thus in analogy to superexchange (described later in this section), this intermediate orbital plays the role of filled orbital in the super-exchange interaction \cite{fazekas1999lecture}.

 However, the impurity orbital in $C_2H$ is not entirely filled (see band structure in Fig. \ref{fig:bschcf}). This opens another root for indirect exchange mechanism called double-exchange \cite{fazekas1999lecture}. So to explain the origin of indirect-exchange mechanisms in $C_2H$ and $C_2F$ materials the super-exchange and double-exchange mechanisms are reviewed.

\subsubsection{Super-Exchange Magnetization Mechanism}
\begin{figure}[H]
\centerline{\includegraphics[height=2cm]{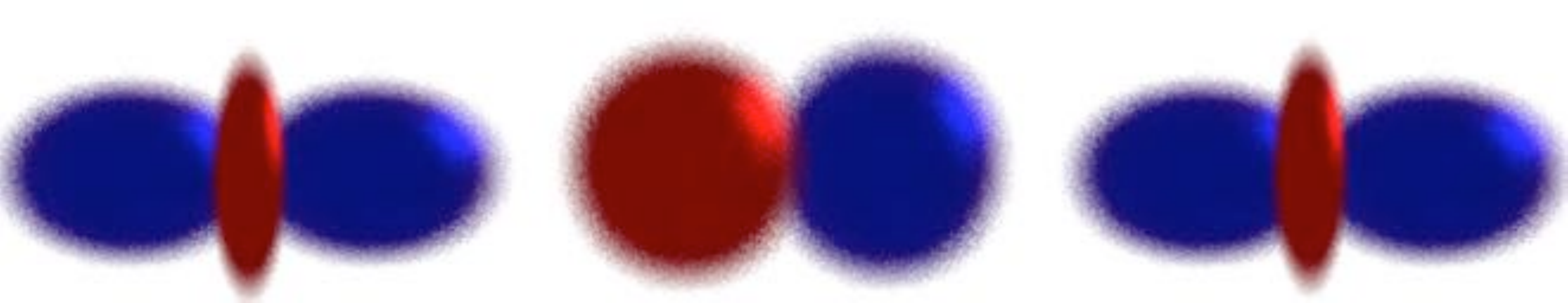}}
\caption{Super-exchange mechanism: Two unfilled $d$-orbitals hybridize with the intermediate $p$-orbital of oxygen in AFM transition metal oxides. } 
\label{fig:super}
\end{figure}

Super-exchange mechanism is an indirect exchange between unfilled orbitals through an intermediate filled orbital. The anti-ferromagnetism (AFM) of most of the transition-metal compounds can not be interpreted by direct exchange. $d$-orbitals are so localized thus the hopping only can happen between orbitals on nearest neighbor atoms. Most of AFM insulators are transition-metal oxides where transition-metal cations separated by the oxygen anions. In these systems, direct hopping between $d$-orbitals is negligible so one needs to extend the direct exchange concept to these cases by considering hopping via intermediate filled orbital of oxygen (see Fig. \ref{fig:super}). In these systems depending on the type of orbitals and the relative angle between the orbitals different magnetic orders can exist. For example, in the case of AFM, two unfilled orbitals hybridize with the same orbital of intermediate state in $180^{\circ} $ arrangement (Fig. \ref{fig:super}). 

In super-exchange mechanism, the intermediate orbital is filled. So usually super-exchange mechanism occurs in insulating materials.

\subsection{Double-Exchange Magnetization Mechanism}
Double-exchange usually encounters in mixed valance compound. In these systems usually the number of electrons per site is not an integer number. So the electrons without spending coulomb cost $U$ can hop between sites. Therefore, double-exchange usually happens for metallic compounds \cite{fazekas1999lecture}. 

\begin{figure}[H]
\centerline{\includegraphics[height=12cm]{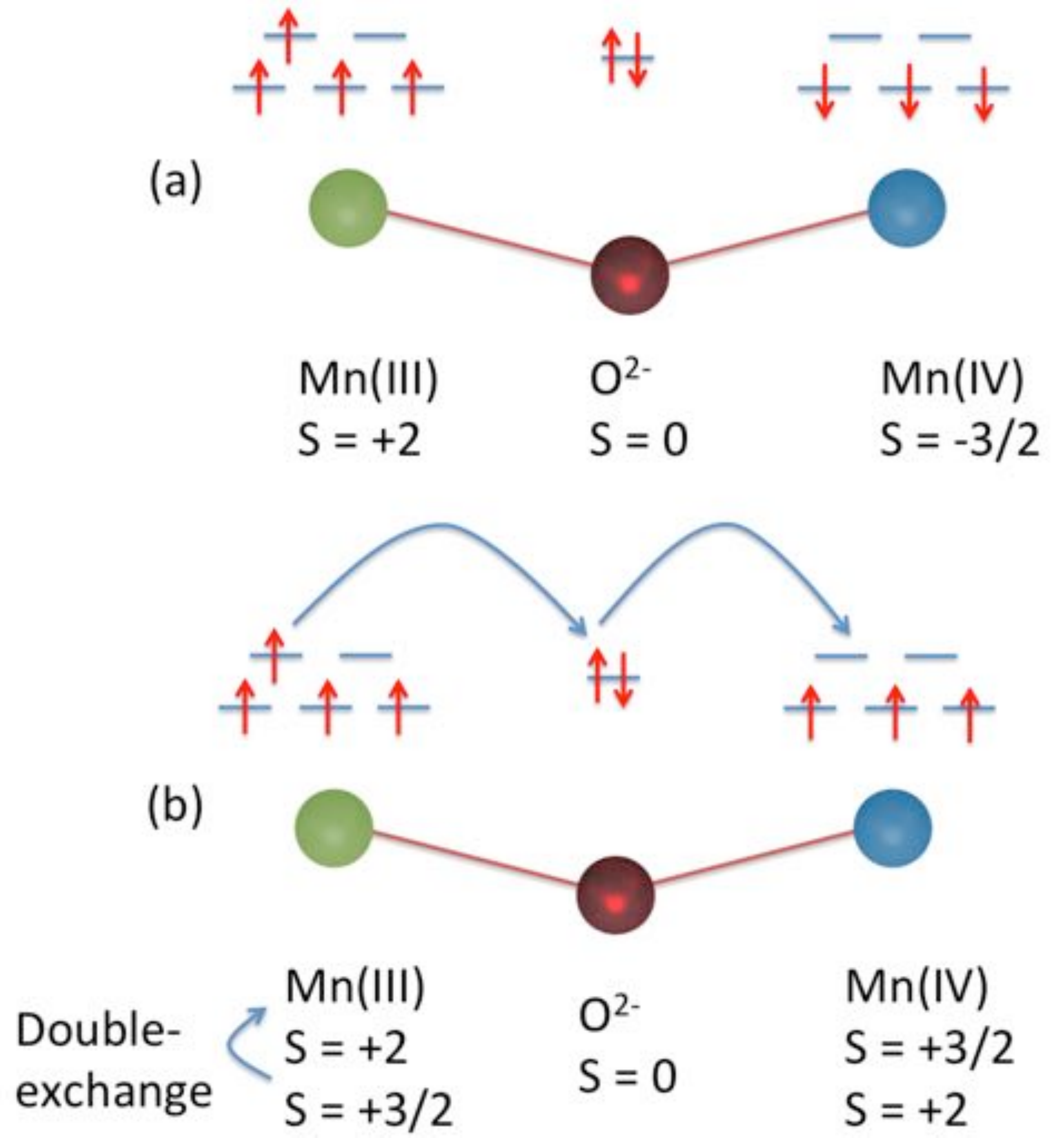}}
\caption{Example of Super-exchange V.S. double-exchange mechanisms: (a) Super-exchange with transferring spin information without charge transfer in Mn(III)-O-Mn(IV) compound. This system favors an AFM order. (b) In Double-exchange, two electrons are simultaneously transferring in which the up electron is hopping from oxygen to Mn(IV) while it is replaced by another electron with spin up hopping from Mn(IV) \cite{powell2010molecular}. } 
\label{fig:double}
\end{figure}

Fig. \ref{fig:double} shows the difference between double-double-exchange and super-exchange mechanisms in mixed valance Mn(III)-O-Mn(IV) compounds. In superexchange, only spin information is exchanged but in double-exchange charge and spin are transferring simultaneously.

\subsection{Indirect-Exchange Mechanisms in Graphone}
In graphone, indirect-exchange is an additional contribution to direct-exchange mechanism. Depending on the type of passivation (H or F), one can see different indirect-exchange mechanisms in half-passivated graphene. The projected band-structure into impurity orbitals in Fig. \ref{fig:bschcf} shows that in $C2F$, $sp$-orbital is entirely filled (red band-diagram). The filled impurity orbital in $C_2F$ is in analogy to oxygen cations in transition-metal oxide compounds (Fig. \ref{fig:double}). Thus the impurity orbital in $C_2F$ is an intermediate site for super-exchange mechanism. 

\begin{figure}[H]
\centering
\includegraphics[width=0.8 \textwidth]{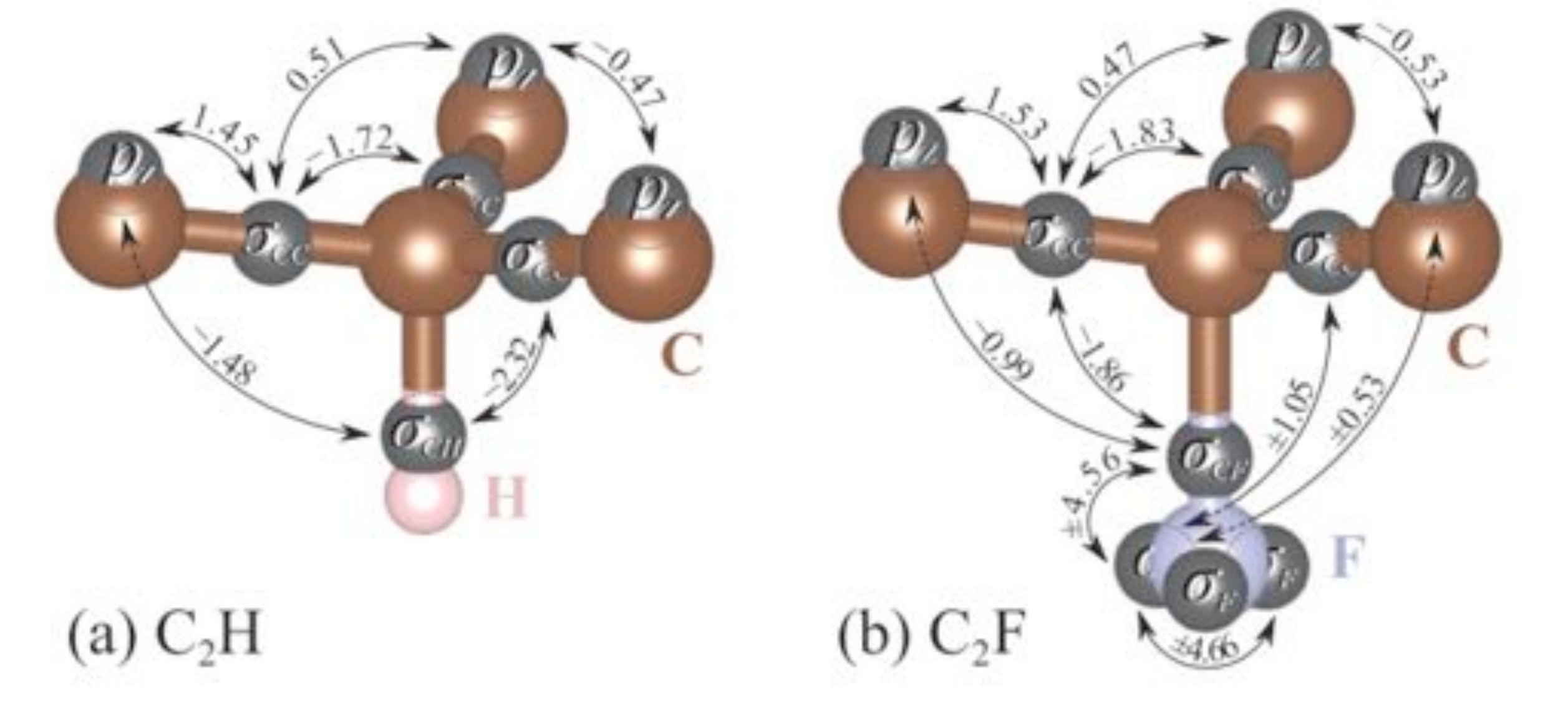}
\caption{The hopping parameters (in eV) between different Wannier orbitals for a) $C_2H$ b) $C_2F$. The sign in hopping parameters show the type of overlap between orbitals, positive for constructive and negative for destructive. Figure is reprinted with permission from Ref. \cite{PhysRevB.88.081405}.} 
\label{fig:hopping}
\end{figure}

However, in the case of $C_2H$, band-structure shows that the impurity orbital is not filled. The hopping parameters between impurity orbital ($s$-orbital) in Fig. \ref{fig:hopping} shows that double-exchange is an additional pathway since the unfilled impurity orbital opens a new channel for charge transfer. Thus spin together with charge can be transferred between unpaired electrons in carbon and impurity orbitals. 

\section{Magnetic Properties of Graphone/h-BN}\label{sec:Magnetic}
So far in this chapter the magnetic properties of pristine graphone have been described. It has been shown by spin-polarized first principle calculation in LSDA scheme that half-hydrogenated graphene (A or B) has about 1 $\mu_{B}$ per unit-cell magnetization \cite{zhou2009ferromagnetism}. Unfortunately, as it has been described in chapter \ref{chap:graphone}, pristine graphone is not stable enough to be of any practical use.  

In the previous section, it has been explained that h-BN substrate considerably increases the graphone(B) stability, making it feasible for further studies. It is, however, important to show that such stabilized graphone still has the magnetic properties expected from the pristine graphone. In this section,  the results of our LSDA with GGA calculations for graphone(B) on h-BN substrate are presented.  

\begin{figure}[H]
\centerline{\includegraphics[height=5cm]{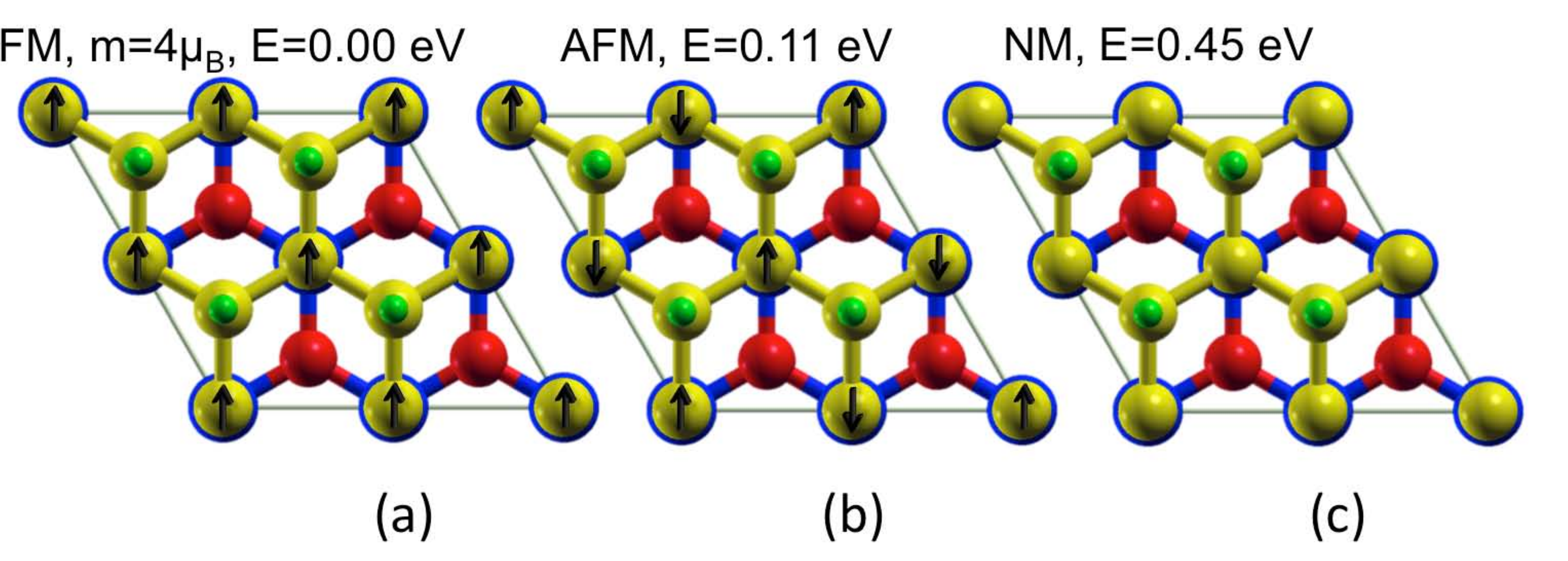}}
\caption{Different magnetization states of graphone(B)/h-BN with their relative energy: a) Ferromagnetic with 1 $\mu_{B}$ per unit-cell magnetization which is the most stable configuration, b) AFM state with zero magnetization, and c) normal state.} 
\label{fig:mag}
\end{figure}

The calculations were perfomed with fully relativistic PBE-GGA pseudo potential with non-linear core corrections and cold smearing with 0.001 eV degauss. Our supercell consists of $2\times2$ unit-cells with 20 $AA$ vacuum layer and $16\times16\times1$ Monkhorst-Pack k-points mesh. The result shows about 1 $\mu_{B}$ magnetization and a 0.42 eV (per unit-cell) energy difference between the ferromagnetic state (FM) and normal state, see Fig. \ref{fig:mag}.

\chapter[SPIN-ORBIT COUPLING IN GRAPHENE]{\uppercase{Spin-Orbit Coupling in Graphone} \footnote[1]{Part of the data reported in this chapter is reprinted with permission from my work in Ref. \cite{PhysRevB.90.035433}} }\label{chap:SOC}
Graphene due to its relatively small spin-orbit coupling (SOC) and large mobility offers various applications in spintronics such as spin transport \cite{tombros2007electronic}, spin qubits \cite{trauzettel2007spin} and spin injection \cite{PhysRevLett.105.167202}. However, weak SOC in graphene which favors spin transport, hampers other application of graphene in spintronics. 

Spin-orbit coupling plays a central role in emerging variety of phenomena such as spin relaxation, spin transport and quantum spin hall effects in spintronics \cite{RevModPhys.76.323, bernevig2006quantum, murakami2006quantum, kane2005z, wolf2001spintronics}. Carbon is a light atom and  the intrinsic SOC in carbon atom is small (12 meV \cite{PhysRevB.74.155426}). Itinerant electrons in graphene mainly belong  to $p_z$-orbitals which has zero angular momentum ($m_l = 0$). So the band-splitting due to the intrinsic SOC in graphene ($2 \lambda_{I} = 24 \mu $ eV)  is even smaller than for example corresponding value in diamond ($\sim meV$). This small value manly arises due to the $p_z-d$ orbital mixing discussed extensively in Ref. \cite{PhysRevB.80.235431, PhysRevB.82.245412}.

\section{Methods to Increase Spin-Orbit Coupling in Graphene}
There are few methods to enhance SOC in graphene. These methods usually utilize enhancement of local electric field induced by the substrate \cite{PhysRevLett.100.107602, PhysRevLett.99.236809, avsar2014spin} or symmetry reduction due to the chemisorbed-impurities \cite{PhysRevLett.110.246602, PhysRevB.91.115141}. 

In the first method the substrate introduces additional source to Rashba SOC in graphene \cite{PhysRevLett.100.107602, PhysRevLett.99.236809, avsar2014spin}. Rashba SOC is induced in the systems with broken inversion symmetry such as crystal interface and surface of the materials with structural inversion asymmetry \cite{maekawa2012spin}. Rashba Hamiltonian can be written as

\begin{eqnarray}\label{eq:rashba}
H_{R} = \alpha_{R} (E \times p).s
\end{eqnarray}
 
where electric field $E$ and momentum $p = \hbar k$  will introduce an effective magnetic field ($E \times p$) perpendicular to $k$. Rashba coupling $\alpha_R$ is material dependent and is proportional to the effective field in addition to  strength of SOC. By placing graphene on the top of substrates such as Ni(111) \cite{PhysRevLett.100.107602} and YIG \cite{PhysRevLett.114.016603} one can enhance the effective electric field at the interface which results in fairly large SOC ($\sim$ 10 meV).

Another method to introduce SOC in graphene is chemisorption of graphene with adatoms. In this case even light adatoms like H and F can induce a fairly large SOC in graphene \cite{PhysRevLett.110.246602, PhysRevB.91.115141}. Chemisorbed adatoms like H and F introduces buckling in graphene layer (e.g. 0.32  in the case of $C_2H$) so gives rise to the change in the original carbon orbital hybridization in graphene ($sp^2$). In these systems the hybridization of carbon changes from planar $sp^2$-type hybridization (with pseudo-spin inversion symmetry) to more like $sp^3$-type with broken pseudo-spin inversion symmetry. This change in hybridization result in a great enhancement in SOC of conduction electrons \cite{PhysRevLett.110.246602}

The induced SOC due to the chemisorption enhances by increasing the buckling \cite{PhysRevLett.110.246602}. It has been shown that SOC energy splitting in the case of fluorinated-graphene can reach $30$ meV \cite{PhysRevB.91.115141}. This value for SOC energy splitting is one order of magnitude greater than Hydrogenated graphene ($\sim$ 1 meV) due to the fluorine orbitals and its hybridization with carbon orbitals. \cite{PhysRevB.91.115141}.

 In Refs. \cite{PhysRevLett.110.246602, PhysRevB.91.115141}, the adatoms effect on SOC has been studied extensively based on the concentration of adatoms. In this section, the effect of passivation together with substrate (h-BN) on inducing spin-orbit coupling in graphene layer has been described. 

\subsection{Spin-Orbit Coupling in Graphone/h-BN}
\begin{figure}[!h]
\centering
\begin{subfigure}[b]{1\textwidth}
\centering
\includegraphics[height=10cm]{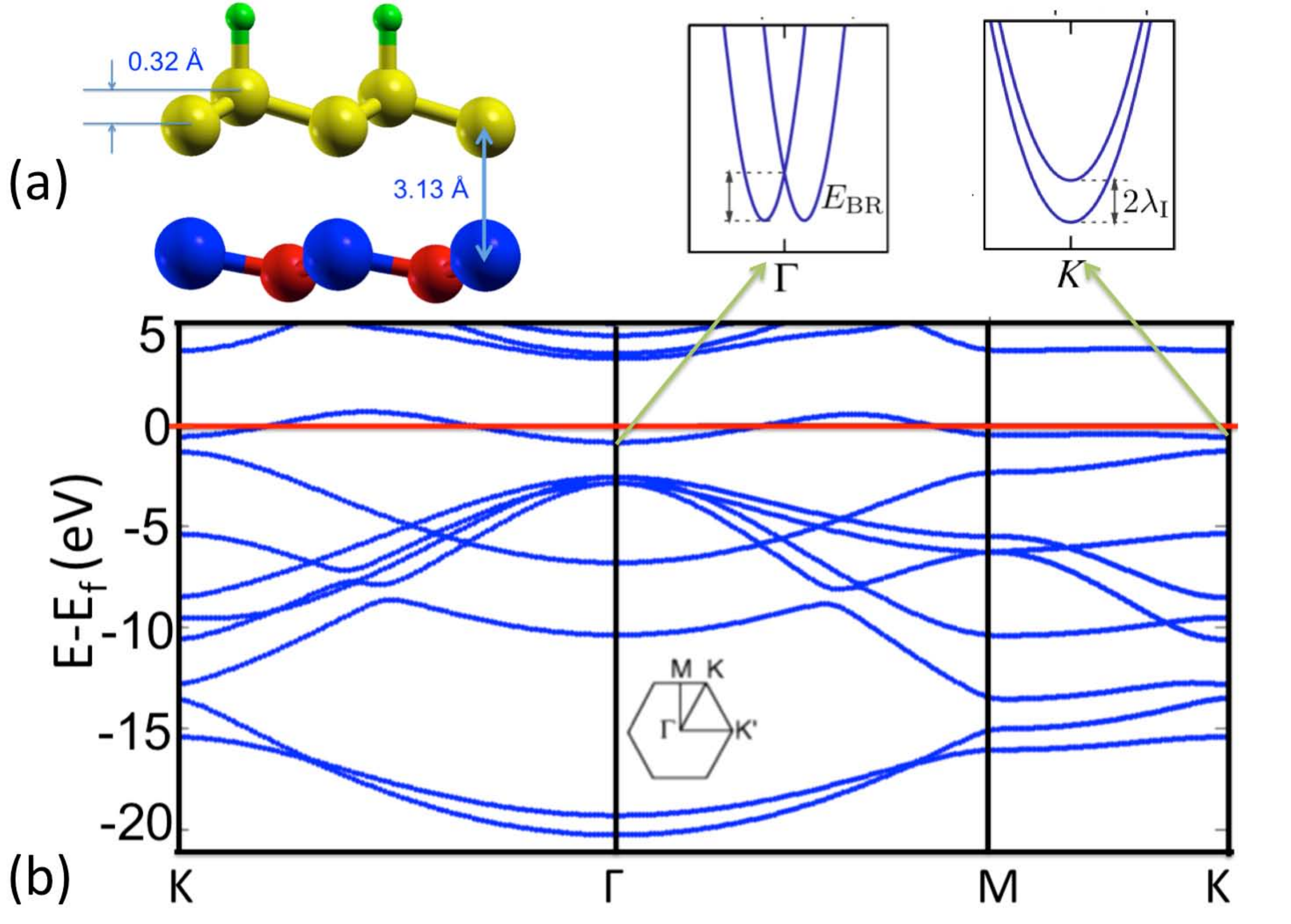}
\caption{}
\end{subfigure}
\caption{a) Optimized structure for graphone(B)/h-BN b) Band-band-structurestructure of graphone/h-BN (fully relativistic GGA) with considering SOC}
\label{fig:bs}
\end{figure}

In this section, the effect of substrate (h-BN) together with passivated on spin-orbit coupling of graphene is investigated. It has been shown hydrogen atoms by changing hybridization of graphene from $sp^2$- to $sp^3$-type hybridization induces a relatively large spin orbit coupling in graphone \cite{PhysRevLett.110.246602}. In this section it is shown that h-BN enhances the Rashba SOC by introducing an additional effective electric field in graphone layer (see equation \ref{eq:rashba}). To understand both passivation and substrate effect on SOC, the band-structure of the optimized graphone(B)/h-BN in the presence of SOC is shown in Fig. \ref{fig:bs}. Group symmetry near K-point and $\Gamma$-point is $C_3$ and $C_{3\nu}$, respectively whereas the  $\pi^{*}$ band is located in the vicinity of the Fermi-energy \cite{PhysRevLett.110.246602}. Up to the first order in momentum near the K-point and  $\Gamma$-point these bands are described by the following SOC Hamiltonians:

\begin{equation}
H_{\mbox{\textsc{soc}}}^{\tau K}=\lambda^{\mbox{\textsc{br}}} (k_{x} s_{y}-k_{y} s_{x}) + \tau \lambda^{\mbox{\textsc{i}}} s_{z} ,
\label{soc}
\end{equation}

\begin{equation}
H_{\mbox{\textsc{soc}}}^{\Gamma}=\lambda^{\mbox{\textsc{br}}} (k_{x} s_{y}-k_{y} s_{x}) .
\label{soc}
\end{equation}

Two terms in equation \eqref{soc}, correspond to induced Bychkov-Rashba-like SOC with $\lambda^{\mbox{\textsc{br}}}=0.14$ meV and intrinsic SOC with $\lambda^{\mbox{\textsc{i}}}=1.24$ meV, respectively. $\tau$ is $+1$  ($-1$) for K (K') point.

$\lambda^{\mbox{\textsc{i}}}$ was obtained as half of the band-splitting of $\pi^{*}$ band in the vicinity of the Fermi-energy at K point ($2.4$meV) and $\lambda^{\mbox{\textsc{br}}}$ is obtained at $\Gamma$ point. This SOC is  comparable to fully hydrogenated graphene in both sides (chair graphane) \cite{zhou2010enhanced}, which has near $8.7$meV band-splitting.

The increase of $\lambda^{\mbox{\textsc{i}}}$ in comparison to $\lambda^{\mbox{\textsc{i}}}=12 \mu eV$ for graphene \cite{PhysRevB.82.245412} is due to the bending of carbon bonds, see Figure \ref{fig:bs}. This bending changes the $sp^{2}$, plane graphene, hybridization to $sp^3$, diamond like, hybridization. This value of $\lambda^{\mbox{\textsc{i}}} $ is comparable to fully hydrogenated-graphene in both sides (chair graphane \cite{zhou2010enhanced}), which reported near 8.7 meV band-splitting.

By comparing $\lambda^{\mbox{\textsc{br}}}$ obtained for grahone/h-BN ($\lambda^{\mbox{\textsc{br}}} $= 0.14 meV) and pristine graphone ($\lambda^{\mbox{\textsc{br}}}\leq $0.1 meV \cite{PhysRevLett.110.246602}) I conclude the effect of substrate (h-BN) on Rashba SOC due to enhancing the effective electric field in graphone layer.

The increase of $\lambda^{\mbox{\textsc{i}}}$ in comparison to $\lambda^{\mbox{\textsc{i}}}=12 \mu eV$ for graphene \cite{PhysRevB.82.245412} is due to the bending of carbon bonds, see Figure \ref{fig:bs}. This bending changes the $sp^{2}$, plane graphene, hybridization to $sp^3$, diamond like, hybridization. This value of $\lambda^{\mbox{\textsc{i}}} $ is comparable to fully hydrogenated graphene in both sides (chair graphane \cite{zhou2010enhanced}), which reported near 8.7 meV band-splitting.
%
%
%

\chapter[GRAPHONE APPLICATION IN SPINTRONICS: GRAPHONE SPIN-VALVE DEVICES]{\uppercase{GRAPHONE APPLICATION IN SPINTRONICS: GRAPHONE SPIN-VALVE DEVICES} \footnote[1]{Part of the data reported in this chapter is reprinted with permission from my work in Ref. \cite{PhysRevB.90.035433}} } \label{chap:TMR}

The tunneling magnetoresistance (TMR) occurs in magnetic tunnel junctions that consist of two ferromagnetic metallic layers separated by a thin insulator (tunneling current). The tunneling current is then controlled by switching magnetization in one of the ferromagnetic electrodes and TMR is a result of spin dependent tunneling.    

\begin{figure}[h]
\centering
\includegraphics[width=0.7\textwidth]{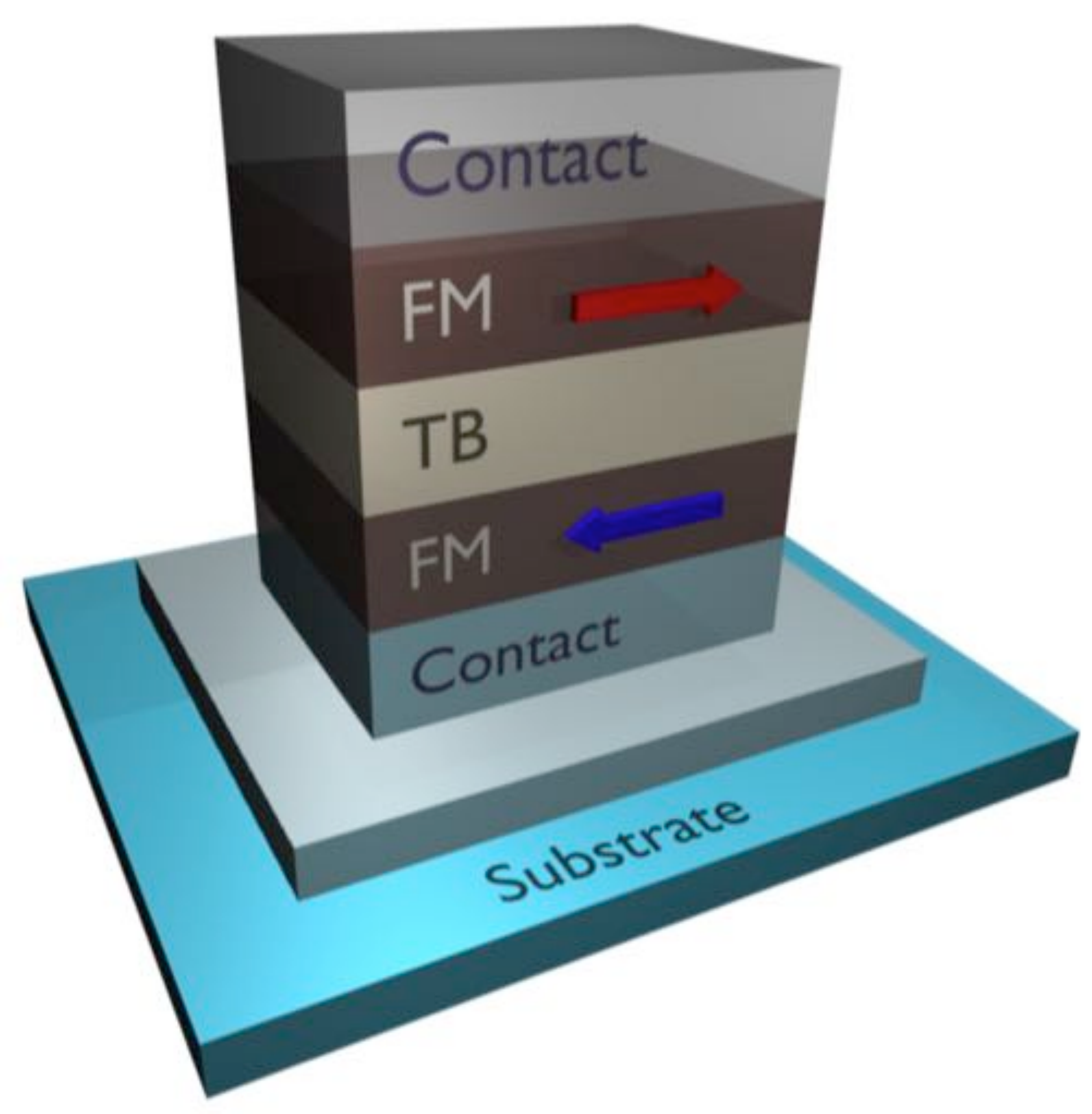}
\caption{Tunneling magnetoresistance device (TMR) consists of two ferromagnetic electrodes and insulating barrier in between and using spin degrees of freedom to control tunneling current.} 
\label{fig:tmr}
\end{figure}

TMR can be identified from Julliere's model with the following assumptions: i) spin of electrons is not changing during the tunneling (two spin channels are decoupled during the tunneling and independent) ii) conductance for each spin channel is proportional to the product of density of states f the two ferromagnetic electrodes. It is important for these devices that the ferromagnetic layers have a high spin polarized electronic state near the fermi energy. Then based on these assumptions one can define TMR ratio as follows:

 \begin{eqnarray}
TMR \equiv \frac{R_{AP} - R_p}{G_{AP}} = \frac{G_{P} -G_{AP}}{G_{P}}{G_P} = \frac{J_{p} - J_{AP}}{J_P}.
\end{eqnarray}

Note according to the second assumption $J^{\alpha} \propto D_{1}^{\alpha}(E_F) D_{2}^{\alpha}(E_F)$ ($\alpha$ is corresponding to two spin channels $\uparrow$ and $\downarrow$). $D_{1}^{\alpha}(E_F)$ and $D_{1}^{\alpha}(E_F)$ are density of states for two ferromagnetic electrodes. 
 
 \begin{eqnarray}
 TMR =1 - \frac{J_{AP}}{J_P} = 1- \frac{J^{\uparrow}_{AP}+J^{\downarrow}_{AP}}{J^{\uparrow}_{P}+J^{\downarrow}_{P}} = \frac{2P_1P_2}{1-P_1P_2},
 \end{eqnarray}
 
where $P_1$ and $P_2$ are spin polarization for two electrodes $P_i = \frac{D_{i}^{\uparrow}(E_F)-D_{i}^{\downarrow}(E_F)}{D_{i}^{\uparrow}(E_F)+D_{i}^{\downarrow}(E_F)}$ (i=1,2 indicates the order of ferromagnetic electrodes).

\begin{figure}[h]
\centering
\includegraphics[width=1\textwidth]{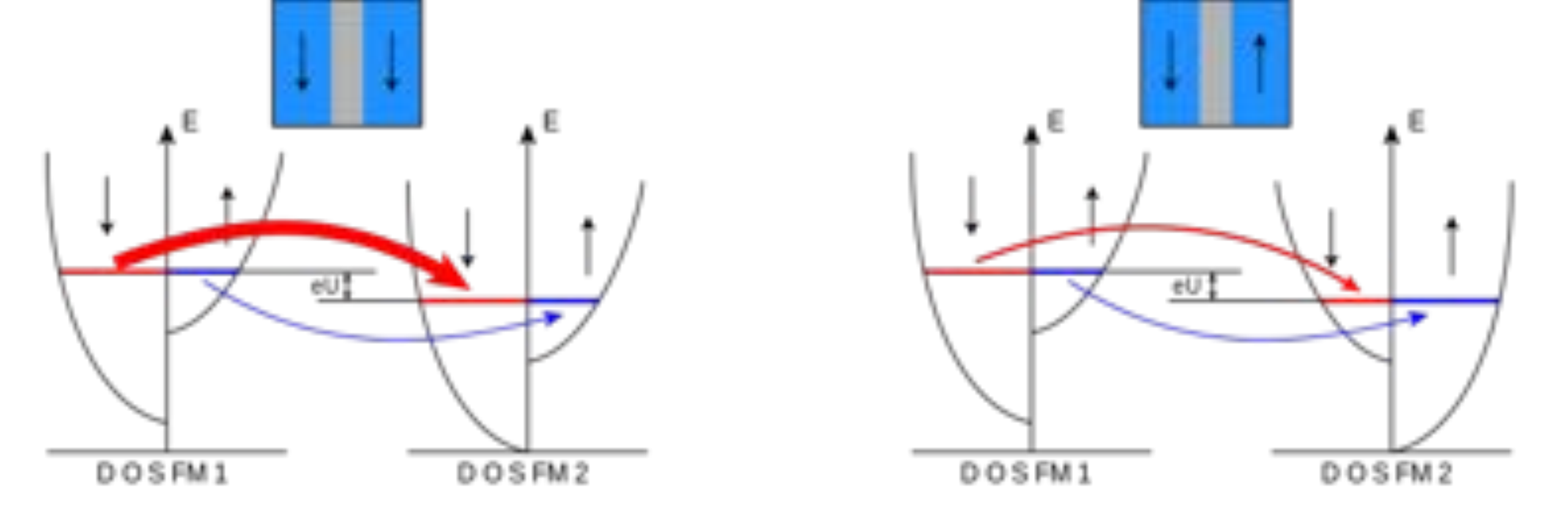}
\caption{Two-current model: For tunneling between two ferromagnets  with parallel (left) and anti-parallel (right) alignment of the magnetizations. Maximum tunneling current occurs in the parallel alignment (left) and minimum in the anti-parallel alignment.} 
\label{fig:tmr}
\end{figure}

\section{Vertical Graphone Tunneling Magnetoresistance (TMR) Heterostructures: Graphone/h-BN/Graphone} \label{sec:vtmr}
In this section,  it has be shown that graphone/h-BN multilayer heterostructures possess nearly 100\% polarized states at the fermi energy and thus are perfect half metals.

We will also see the magnetic properties of the graphone can be controlled and enhanced by changing the number of layers of h-BN. h-BN has a large band gap (6 eV) and can be a good insulator for fabrication of TMR devices. These properties of the multilayer graphene h-BN heterostructures can open new horizons in spintronics.

I have performed the same LDA, GGA, and VDW structure calculation for two layers of graphone(B) with 1, 2, and 3 layers of AA-stacked h-BN sandwiched in between. The results are shown in Table \ref{table:2}. 

\begin{table}[!h]
\centering
\begin{tabular}{|c|cc|cc|cc|}
\hline 
Structure & LDA& & GGA & & VDW & \\
&$d_{CB}$ & $d_{CC}^{\perp}$ &$ d_{CB} $&$ d_{CC}^{\perp}$ &$ d_{CB}$ &$ d_{CC}^{\perp}$ \\ 
\hline
graphone & 3.12 & 0.38 & 3.13 & 0.30 & 3.13 & 0.32 \\ 
g/1bn/g & 2.60 & 0.28 & 2.72 & 0.29 & 3.19 & 0.30 \\ 
g/2bn/g & 1.95 & 0.38 & 2.10 & 0.38 & 3.08 & 0.30 \\ 
g/3bn/g & 2.48 & 0.28 & 2.47 & 0.30 & 3.01& 0.31\\ 
\hline
\end{tabular} 
\caption{Optimized structure (in \AA): $d_{CB}$ is the distance between carbon on top of the boron atom and $d^{\perp}_{cc}$ is the vertical distance between two carbon atoms which belong to the same layer. It shows the magnitude of the bond bending.} 
\label{table:2}
\end{table}

\begin{figure}[H]  
\hspace{1mm}
\raisebox{+0.5\height}{\includegraphics[height=3.cm, scale=0.1]{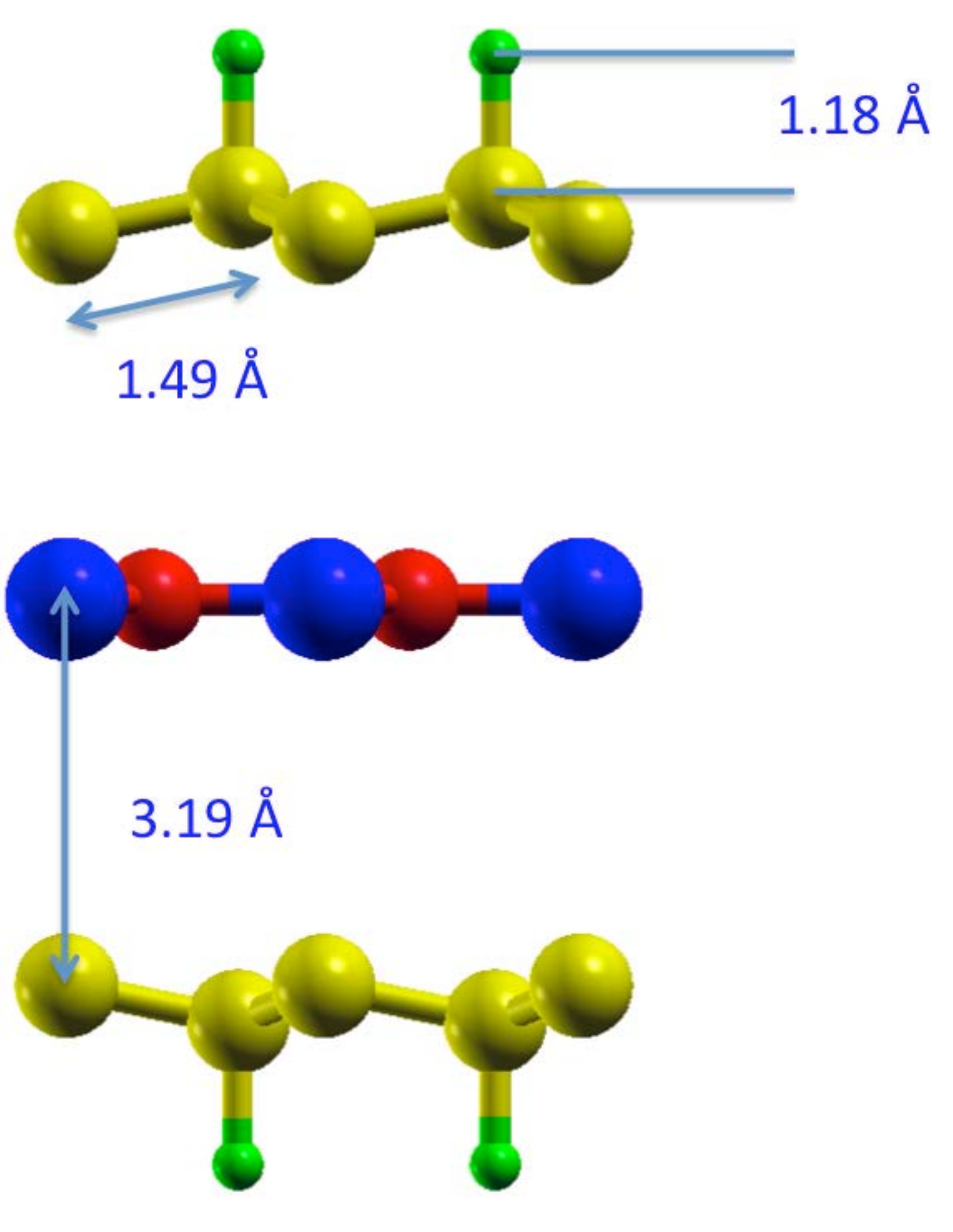}} 
\hspace{5mm}
\raisebox{+0.1\height}{\includegraphics[height=6cm, width= 11.8cm]{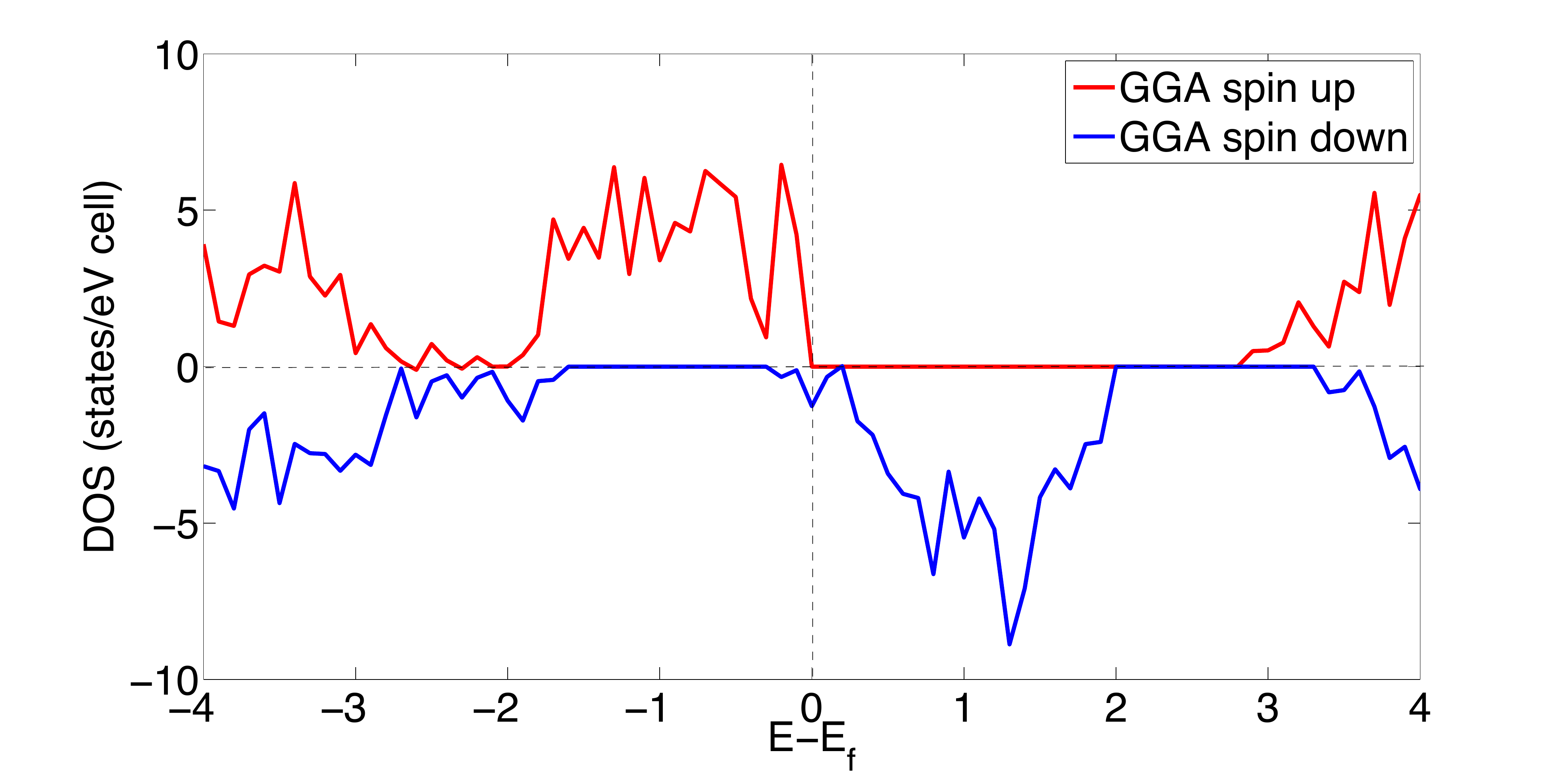}}
\raisebox{+0.5\height}{\includegraphics[height=4cm, scale=0.16]{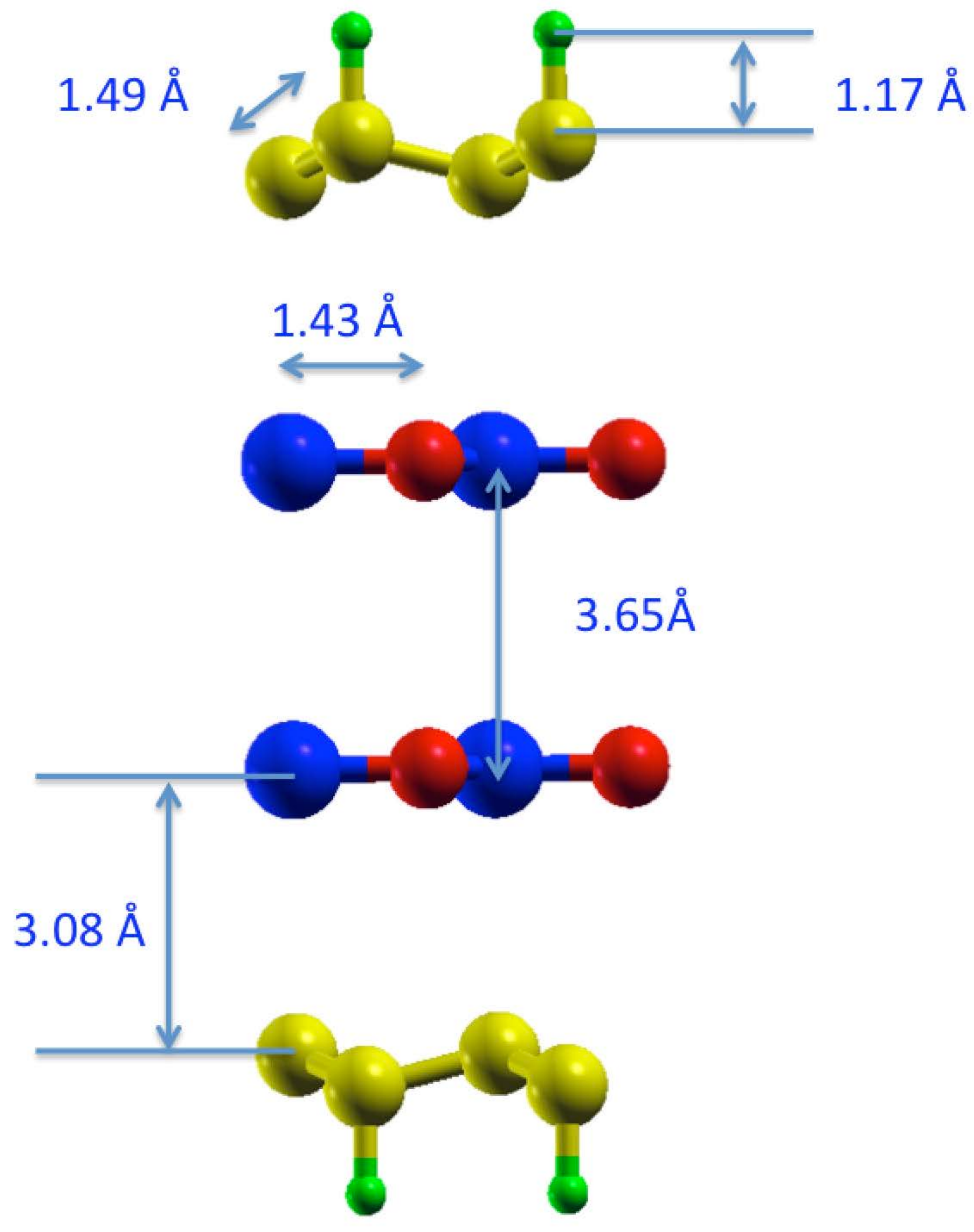}}
\raisebox{+0.1\height}{\includegraphics[height=6cm, width= 11.8cm]{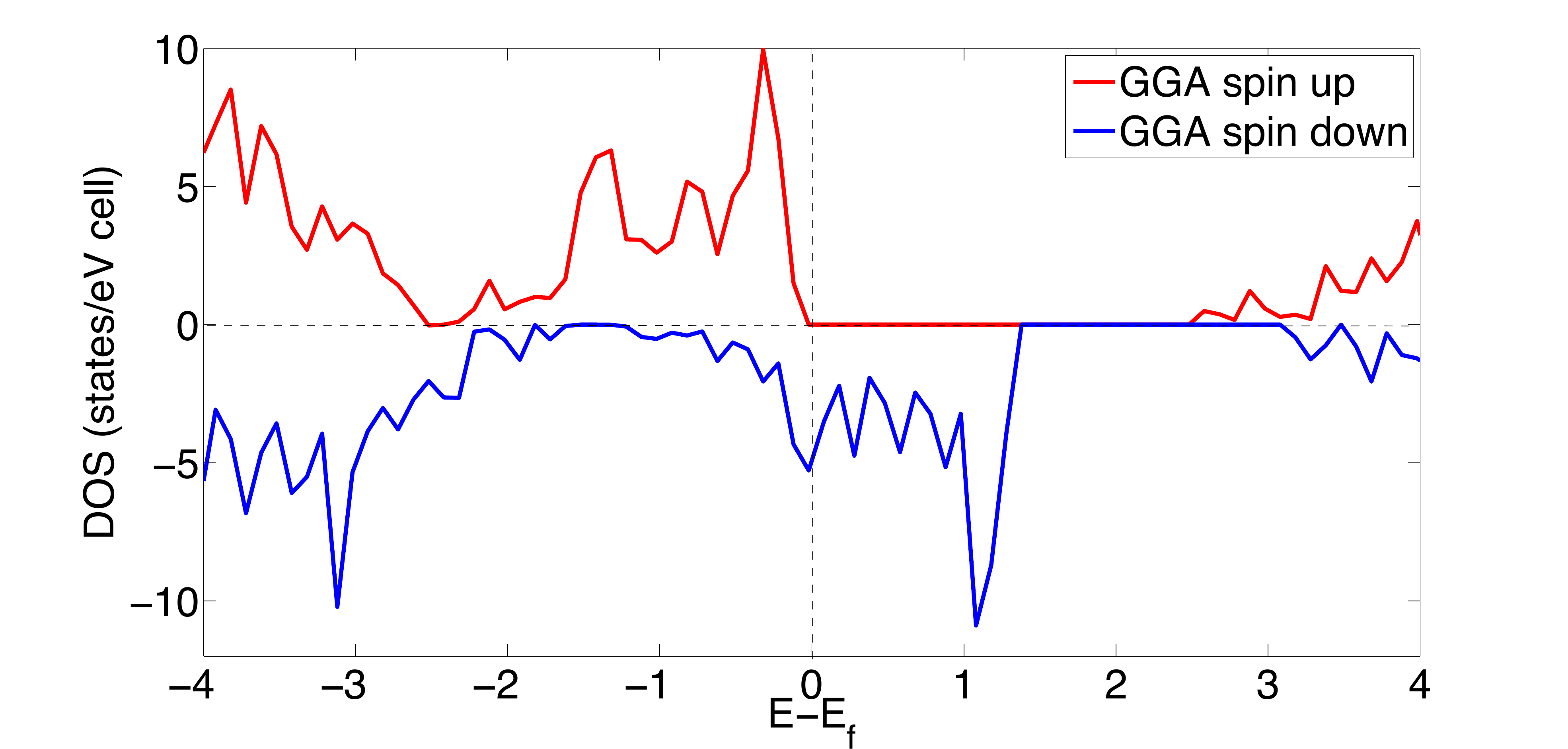}}\hspace{7mm}
\raisebox{+0.2\height}{\includegraphics[height=5cm, scale=0.18]{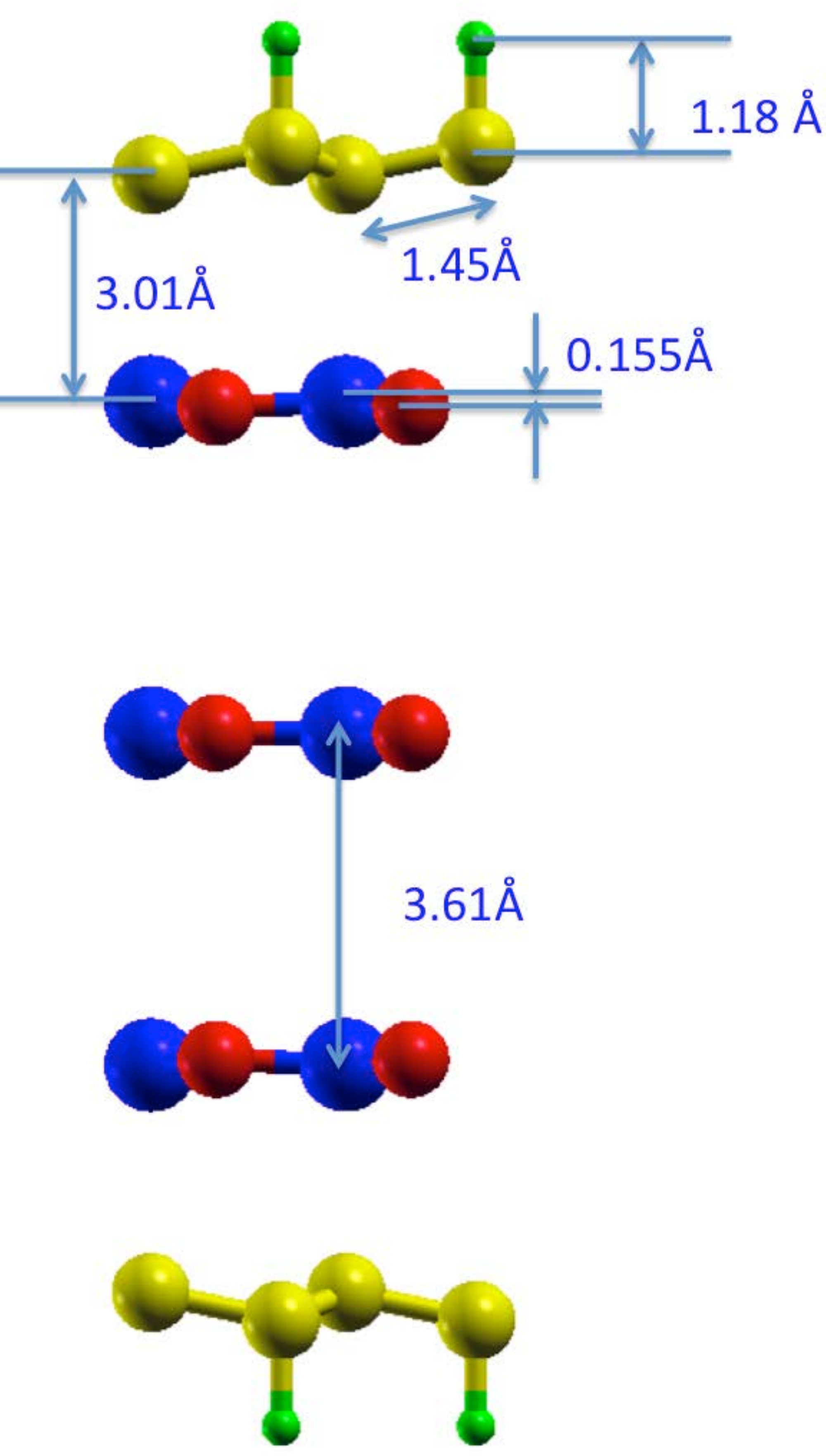}}
\hspace{4mm}
\raisebox{+0.04\height}{\includegraphics[height=6cm, width= 11.6cm]{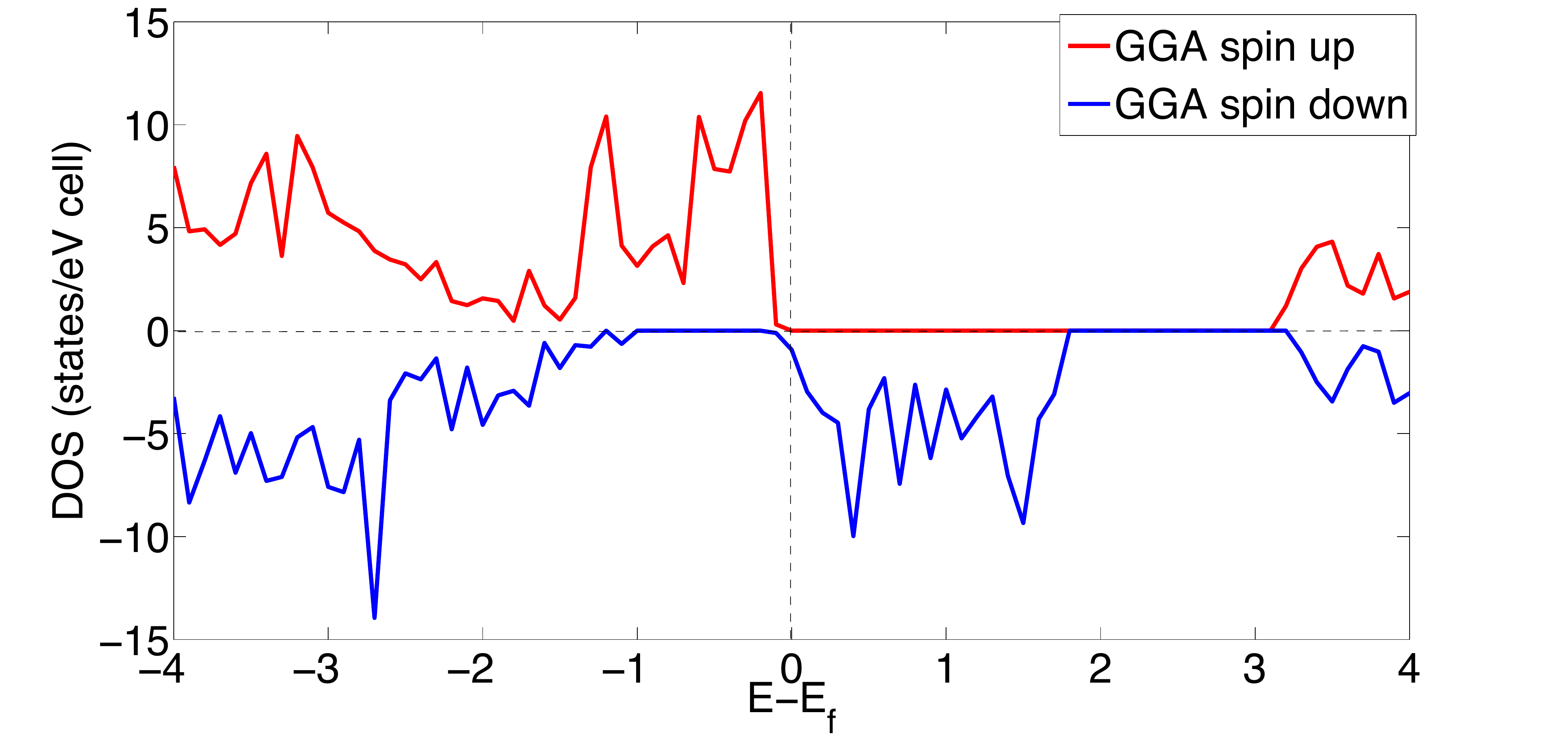}}\hspace{7mm}
\caption{Optimized structure and spin-polarized DOS (all the energies are in eV) for double-layered graphone(B) and a) single , b) double and c) triple-layered AA-stacked h-BN in between.}
\label{fig:tmrdos}
\end{figure}

I have calculated the spin polarized density of states for the three multilayer structures using LSDA scheme with 0.01 eV degauss and cold smearing. The structural parameters for this calculations have been taken from the VDW results. The results are shown in Fig. \ref{fig:tmrdos}.

All three heterostructures show half metallic behavior near the fermi-energy. 
The spin up gap varies between 1 and 2 eV. According to Julliere model \cite{julliere1975tunneling}, such half metallic materials with 100$\%$ spin polarization are ideal for fabrication of TMR devises.
\FloatBarrier

\section{Lateral Graphone Tunneling Magnetoresistance (TMR) Heterostructures: Graphone-Graphane-Graphone}
In section \ref{sec:vtmr} vertical graphone/h-BN/graphone has been introduced. However, stability of two graphone electrodes on both sides in one hand and integration of these vertical TMR devices in another hand are two main challenges in front of fabrication of the proposed vertical TMR devices from graphone/h-BN heterostructures. 

One solution is replacing one of the electrodes with other ferromagnets like Ni and Co. This method breaks the symmetry in the heterostructure and may cause some interface issues due to the coupling of h-BN with other ferromagnet.

\begin{figure}[!h]
\centering
\includegraphics[width=1\textwidth]{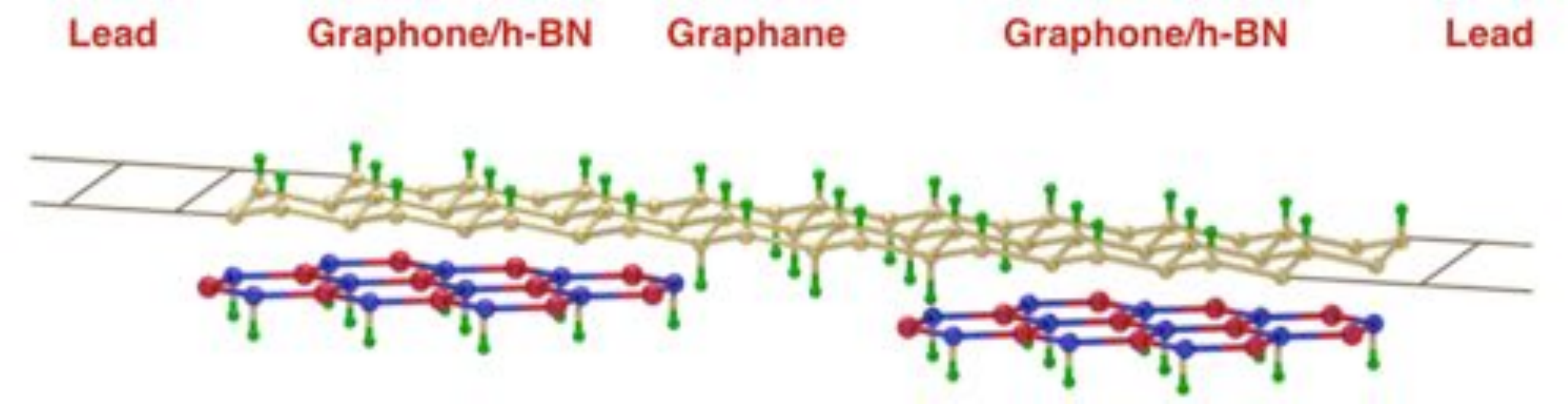}
\caption{Proposed lateral graphone-graphane-graphoneTunneling magnetoresistance device (TMR) consists of two ferromagnetic graphone electrodes (on top of h-BN to stabilize graphone as in proposed in section  \ref{sec:vtmr}) and graphane as an insulating barrier in between. } 
\label{fig:ltmr}
\end{figure}

Another solution to the problem is implementing graphone in a lateral TMR heterostructure (see Fig. \ref{fig:ltmr}) \cite{hemmatiyan2015lateral}. In this system, two ferromagnetic graphone are separated by fully hydrogenated graphene on both sides (Graphane) as an insulating barrier. The proposed experimental set up is as follows: substrate effect of h-BN helps to control passivation as it discussed in chapter \ref{chap:graphone}). Starting from graphene/h-BN,  one can control graphene/h-BN domain sizes by etching part of h-BN (using argon ions) as it proposed in Ref. \cite{liu2013plane}). Then by exposing the entire system to hydrogen plasma, both graphone and graphane are stabilized simultaneously in different regions. Substrate effect of h-BN stabilizes graphone in the graphene/h-BN domains. However, on the etched area (interfaces between graphene/h-BN domains), hydrogen atoms cover graphene from both sides on and result in fully hydrogenated graphene (graphone) \cite{hemmatiyan2015lateral}. 
%

\chapter[SUMMARY AND CONCLUSIONS]{\uppercase{SUMMARY AND CONCLUSIONS}
\footnote[1]{Part of the data reported in this chapter is reprinted with permission from my work in Ref. \cite{PhysRevB.90.035433}}} \label{chap:summary}
In this chapter, I provide a summary of the main results in this thesis. 

\section{Summary and Conclusions}
I propose a feasible way to fabricate graphone in experiment by exposing the graphene/h-BN bilayer to the hydrogen plasma. The dipole moments induced by h-BN orchestrates the hydrogen pattern on the graphene layer. From first principle calculations, I have shown the presence of the preference site for hydrogen adsorption and an increment in migration barrier due to the screening effect of h-BN. The results show induced dipole moments by h-BN will trap hydrogen atoms to only one sublattice and will also kinematically stabilize the graphone layer. 

I also show that the substrate effect of h-BN provides a possible path for fabrication of half-fluorinated graphene in a single sublattice. The substrate prevents fluorine adsorption from both sides. In addition, h-BN reduces distortion of graphene induced during fluoridation thus results in half-fluorinated graphene in single sublattice. 

The calculated band-structures for the optimized graphone(B) on h-BN shows that the screening effect of h-BN not only reduces the band-gap (from near 3 eV in pristine graphone to 1.93 eV in the graphone/h-BN heterostructure \cite{kharche2011quasiparticle}), but also effectively changes the Fermi-energy in the graphone layer.

I calculate the substrate effect in addition to the impurity effect on spin orbit coupling (SOC). I show the chemisorbed  adatoms like H and F enhances intrinsic SOC (by reducing the symmetry). However, h-BN as a substrate enhances local effective electric field on the carbon orbitals results an increment in Rashba SOC.

Finally, I show that multilayer heterostructures (several layers of h-BN are sandwiched in between two layers of graphone) are half metal with near 100\% spin polarization. I propose to use such heterostructures in TMR devices.

I note, that it is also possible to use different elements such as fluorine as an adsorbent on graphene/h-BN heterostructures. These new materials will have different binding energies, migration barriers, and electronic properties. Doping h-BN will change the fermi energy of the graphone layer in addition to affecting hydrogen pattern on the graphene lattice. Therefore, it may provide the mechanism for control of the hydrogen pattern and electronic properties of the heterostructures.


%
%
%

\phantomsection
\addcontentsline{toc}{chapter}{BIBLIOGRAPHY}

\renewcommand{\bibname}{{\normalsize\rm BIBLIOGRAPHY}}

\bibliography{ref}
\bibliographystyle{unsrt}


\end{document}